\documentclass[twocolumn,showpacs,amsmath,amssymb,floatfix,pre,floatfix]{revtex4} 

\usepackage{graphicx,color}
\definecolor{brown}{rgb}{0.63,0.27,0.18}
\definecolor{orange}{rgb}{1.00,0.65,0.00}

\usepackage{moresize}
\usepackage{dcolumn}
\usepackage{bm,multirow}

\begin{document}
\newcommand {\rsq}[1]{\langle R^2 (#1)\rangle}
\newcommand {\rsqL}{\langle R^2 (L) \rangle}
\newcommand {\rsqbp}{\langle R^2 (N_{bp}) \rangle}
\newcommand {\Nbp}{N_{bp}}
\newcommand {\etal}{{\em et al.}}
\newcommand{\Ham}{{\cal H}}
\newcommand{\RalfNew}[1]{\textcolor{red}{#1}}
\newcommand{\scs}{\ssmall}



\title{Density effects in entangled solutions of linear and ring polymers}

\author{Negar Nahali}
\email{nenahali@sissa.it}
\affiliation{SISSA - Scuola Internazionale Superiore di Studi Avanzati, Via Bonomea 265, 34136 Trieste (Italy)}
\author{Angelo Rosa}
\email{anrosa@sissa.it}
\affiliation{SISSA - Scuola Internazionale Superiore di Studi Avanzati, Via Bonomea 265, 34136 Trieste (Italy)}

\date{\today}
\begin{abstract}
In this paper, we employ Molecular Dynamics computer simulations to study and compare the statics and dynamics of linear and circular (ring) polymer chains
in entangled solutions of different densities.
While we confirm that linear chain conformations obey Gaussian statistics at all densities, rings tend to crumple becoming more and more compact as density increases.
Conversely, contact frequencies between chain monomers are shown to depend on solution density for both chain topologies.
Relaxation of chains at equilibrium is also shown to depend on topology, with ring polymers relaxing faster than their linear counterparts.
Finally, we discuss the local viscoelastic properties of the solutions by showing that diffusion of dispersed colloid-like particles is markedly faster in the rings case.
\end{abstract}

\maketitle

\section{Introduction}\label{sec:Intro}
Since the pioneering works of Flory~\cite{Flory1969}, De Gennes~\cite{DeGennes} and Edwards~\cite{DoiEdwards},
excluded volume effects and topological constraints have been recognised playing a fundamental role
in the comprehension of the structural and dynamical properties of polymers in semi-dilute solutions and melts.
Topological constraints (or, {\it entanglements}) hinder the thermal motion of polymers in a way akin to the process of threading a rope out of a pool:
polymer chains can not pass through each other,
while they are allowed to slide past each other.
This fundamental mechanism is recognised to be responsible for the unique properties of polymer solutions and melts~\cite{Flory1969,DeGennes,DoiEdwards,RubinsteinColby}.
Since the very early days of polymer physics,
theoretical speculation and experimental investigation have mainly dealt with solutions and melts of {\it linear} polymer chains~\cite{Flory1969,DeGennes,DoiEdwards,RubinsteinColby}.

More recently a new class of systems has emerged and started receiving similar systematic attention:
entangled solutions and melts of unconcatenated and unknotted circular (ring) polymers~\cite{KhokhlovNechaev85,grosbergJPhysFrance1988,ORD_PRL1994,BreretonVilgis1995,mullerPRE1996,mullerPRE2000,kapnistos2008,Vettorel2009,DeguchiJCP2009,Halverson2011_1,TubianaRosa2011,HalversonPRL2012,SakauePRL2012,Grosberg_PolSciC_2012,SmrekGrosberg2013,RosaEveraersPRL2014,GrosbergSoftMatter2014,CherstvyMetzler2014,CaponeLikosSoftMatter2014,MichielettoTurner2014}.
Why are ring polymers so special and, as we will shortly see, challenging?
According to current theoretical understanding, linear polymers in entangled solutions follow (quasi) ideal statistics because of screening of excluded volume~\cite{DoiEdwards} at large scales,
while ring polymers do not display such an analogous ``compensation'' mechanism and tend to crumple into compact configurations~\cite{mullerPRE1996,mullerPRE2000,Vettorel2009,DeguchiJCP2009,Halverson2011_1,HalversonPRL2012,RosaEveraersPRL2014}.
At odds with their linear counterparts then, topological constraints affect not only the dynamical properties of rings in solution
but they also have consequences on their properties at equilibrium.
Last but not least, the physical behaviour of ring polymers in solution has revealed intriguing connections to the experimentally observed behaviour of chromosomes inside the nuclei of the cells~\cite{Grosberg_PolSciC_2012,RosaEveraersPRL2014}.
In spite of the considerable theoretical and experimental work which have been already dedicated to the subject,
there are still several features of ring polymers in solution which wait to be examined in more detail.

In this article, we present a systematic numerical investigation concerning the equilibrium and dynamical properties of ring polymers in solution,
either as a function of ring size or solution density.
The latter aspect, in particular, has received less attention in the past and this work intends primarily to start filling this gap.
For comparison, we also discuss the same properties for corresponding entangled solutions of linear chains.
As a result, we confirm that while linear chains statistics remains {\it nearly} Gaussian in the long-chain limit {\it at all densities}, rings tend to crumple becoming more and more compact as density increases.
Nonetheless, and interestingly, we prove that increasing the density has also non negligible, measurable effects for contact frequencies in linear chains.
We discuss polymer dynamics, showing that chains and rings of the same size have different relaxation times, with rings relaxing faster.
Then, we conclude the work by addressing the issue of how chain structure affects diffusion of non-sticky colloid particles.

The paper is structured as follows:
In Section~\ref{sec:ModMethods}, we describe the details of the computational methods employed in this work.
In Sec.~\ref{sec:Results}, we present and discuss the results of our work.
Finally, in Sec.~\ref{sec:Concls} we outline the conclusions.

\section{Model and Methods}\label{sec:ModMethods}

\subsection{The model}\label{sec:Model}

\subsubsection{Polymer model}\label{sec:PolymModel}

Linear and ring polymers in solution are described according to the Kremer and Grest~\cite{KremerGrestJCP1990} polymer model.

Excluded volume interactions between beads (including consecutive ones along the contour of the chains) are modelled by the shifted and truncated Lennard-Jones (LJ) potential:
\begin{equation}\label{eq:LJ}
U_{\mathrm{LJ}}(r) = \left\{
\begin{array}{lr}
4 \epsilon \left[ \left(\frac{\sigma}{r}\right)^{12} - \left(\frac{\sigma}{r}\right)^6 + \frac14 \right] & \, r \le r_c \\
0 & \, r > r_c
\end{array} \right. \, ,
\end{equation}
where $r$ denotes the separation between the bead centers. The cutoff distance $r_c=2^{1/6}\sigma$ is chosen so that 
only the repulsion part of the Lennard-Jones is used. The energy scale is set by $\epsilon=k_BT$ and the length scale 
by $\sigma$, both of which are set to unity in our simulations.
Consistent with that, in this work all quantities will be reported in these reduced LJ units.

Nearest-neighbour monomers along the contour of the chains are connected by the finitely extensible nonlinear elastic (FENE) potential, given by:
\begin{equation}\label{eq:Ufene}
U_{\mathrm{FENE}}(r) = \left\{
\begin{array}{lcl}
-0.5kR_0^2 \ln\left(1-(r / R_0)^2\right) & \ r\le R_0 \\ \infty & \
r> R_0 &
\end{array} \right. \, ,
\end{equation}
where $k = 30\epsilon/\sigma^2$ is the spring constant and $R_{0}=1.5\sigma$ is the maximum extension of the elastic FENE bond.

In order to maximize mutual chain interpenetration at relatively moderate chain length~\cite{mullerPRE2000} and hence reduce the computational effort,
we have introduced an additional bending energy penalty between
consecutive triplets of neighbouring beads along the chain in order to control polymer stiffness:
\begin{equation}\label{eq:Ubend}
U_{\mathrm{bend}}(\theta) = k_\theta \left(1-\cos \theta \strut\right) \, .
\end{equation}
Here, $\theta$
is the angle formed between adjacent bonds and $k_\theta = 5 \, k_B T$ is the bending constant.
With this choice, the polymer is equivalent to a worm-like chain with Kuhn length $l_K$ equal to $10 \sigma$~\cite{AuhlJCP2003}.

\subsubsection{Model for colloid particles}\label{sec:ColloidModel}

Colloid-monomer and colloid-colloid interactions are described by the model potentials introduced by Everaers and Ejtehadi~\cite{EveraersEjtehadi2003}.

The total interaction energy between colloid particles at center-to-center distance $r$ can be represented as the sum of two functions:
\begin{equation}\label{eq:Uparticle}
U_{\mathrm{cc}}(r) = \left\{
\begin{array}{lcl}
U_{\mathrm{cc}}^{A}(r) +U_{\mathrm{cc}}^{R}(r) & r \le r_{cc} \\
0 & r > r_{\mathrm{cc}} &
\end{array} \right. \, .
\end{equation}
$U_{\mathrm{cc}}^A(r)$ is the attractive component and it is given by:
\begin{equation}\label{eq:Uapp}
U_{\mathrm{cc}}^{A}(r) = - \frac{A_{\mathrm{cc}}}{6} \left[ \frac{2 a^2}{r^2 - 4 a^2} + \frac{2 a^2}{r^2} + \ln\left( \frac{r^2-4 a^2}{r^2} \right)\right] \, .
\end{equation}
The repulsive component of the interaction, $U_{\mathrm{pp}} ^{R}(r)$, is:
\begin{eqnarray}\label{eq:Urpp}
U_{\mathrm{cc}}^{R}(r)
&=& \frac{A_{\mathrm{cc}}}{37800} \frac{\sigma^6}{r} \left[ \frac{r^2 - 14 a r + 54 a^2}{(r - 2a)^7} \right. \nonumber\\
& & \left. + \frac{r^2 + 14 a r + 54 a^2}{(r + 2a)^7} - 2 \frac{r^2 - 30 a^2}{r^7} \right] \, .
\end{eqnarray}
In previous formulas, $a = 2.5\sigma$ is the colloid particle radius and the constant $A_{\mathrm{cc}}$ = 39.478 $k_{B}T$~\cite{EveraersEjtehadi2003}.
As we are interested to model hard-sphere colloids, the interaction is truncated at its minimum value of $5.595 \sigma$.

Finally, the interaction energy, $U_{mc}$, between a single monomer and a colloid particle with center-to-center distance $r$ is given by:
\begin{equation}\label{eq:Ubp}
\footnotesize{
U_{\mathrm{mc}}(r) = \left\{
\begin{array}{lr}
\frac{2 a^3 \sigma^3 A_{\mathrm{mc}}}{9 (a^2-r^2)^3} \left[ 1 - \frac{(5a^6 + 45 a^4 r^2 + 63 a^2 r^4 + 15 r^6)\sigma^6}{15 (a-r)^6 (a+r)^6} \right] & r \le r_{\mathrm{mc}} \\
 \\
0 & r\ge r_{\mathrm{mc}} \\
\end{array} \right.
}
\end{equation}
where $A_{\mathrm{mc}} = 75.358 k_{B}T$~\cite{EveraersEjtehadi2003}.
Again, since we model non-sticky particles, the potential is cut off at its minimum, $r_{\mathrm{mc}}=3.363\sigma$.

\subsection{Simulation details}\label{sec:SimDetails}
Here, we consider polymer solutions consisting of $M = 160, 80, 40$ circular (ring) or linear chains made of, respectively, $N=250, 500, 1000$ beads each.
The total number of monomers is then fixed to $40'000$ units.
Each polymer solution includes also $100$ colloid particles of diameter $=5\sigma$.
The volume of the simulation box accessible to chain monomers has been chosen in order to fix the monomer density $\rho$
to the values $\rho \sigma^3 = 0.1, 0.2, 0.3$ and $0.4$.

The static and kinetic properties of chains and colloids are studied using fixed-volume and constant-temperature 
molecular dynamics simulations (NVT ensemble) with implicit solvent and periodic boundary conditions.
MD simulations are performed using the LAMMPS package~\cite{lammps}.
The equations of motion are integrated using a velocity Verlet algorithm, in which all beads are weakly coupled to a Langevin heat bath 
with a local damping constant $\Gamma = 0.5 \tau_{\mathrm{MD}}^{-1}$ where $\tau_{\mathrm{MD}} = \sigma(m / \epsilon)^{1/2}$ is the Lennard-Jones
time and $m=1$ is the conventional mass unit for monomer and colloid particles.
The integration time step is set to $\Delta t = 0.012 \tau_{\mathrm{MD}}$.

\subsection{Preparation of initial configurations}\label{sec:IniConfig}
{\it Linear polymers} --
Solutions of {\it linear} chains and colloid particles were prepared first at $\rho\sigma^3=0.1$.
Linear chains were arranged as random walks in space and placed at random positions inside the simulation box.
Random positions were also chosen for colloid particles.
In order to remove possible overlaps between chain monomers and between those and colloid particles,
a short (of the order of a few $\tau_{\mathrm{MD}}$'s) MD run with capped, soft ({\it i.e.} non-diverging) repulsive interactions between monomers and between monomers and colloids was used.

{\it Ring polymers} --
This first setup is not suitable for ring polymers which needs to satisfy the supplementary constraint of avoiding mutual concatenation.
Hence, ring polymers were initially arranged in a very large simulation box, {\it i.e.} at very dilute conditions.
In order to reach the correct monomer density of $\rho\sigma^3=0.1$ we performed then a short (about $400 \tau_{\mathrm{MD}}$ MD steps) simulation by imposing an external pressure on the system which shrinks the simulation box until it reaches the desired value.

For both systems of linear and circular chains,
higher densities were obtained by compressing the solutions by means of higher external pressures.
Of course, during this preparatory phase the complete set of interaction terms described in Sec.~\ref{sec:Model} was employed.

\subsection{Equilibration}\label{sec:MDeq}
Once the system was prepared at the chosen density, we switched to the NVT ensemble.
Polymer solutions with chain sizes of $N=250$, $N=500$ and $N=1000$ beads
were simulated for run times of $12 \times 10^6 \tau_{\mathrm{MD}}$, $24 \times 10^6 \tau_{\mathrm{MD}}$ and $48 \times 10^6 \tau_{\mathrm{MD}}$, respectively.
Unless otherwise stated, chain properties at equilibrium were always calculated on the last tenth of the corresponding trajectory.

In order to check for system equilibration, we have monitored~\cite{KremerGrestJCP1990}
the mean-square displacement of chain monomers in the absolute frame ($g_1(\tau)$) and in the frame relative to the chain centre of mass ($g_2(\tau)$),
and the mean-square displacement of the chain centre of mass ($g_3(\tau)$) at lag-time $\tau$.
They are defined, respectively, as:
\begin{eqnarray}\label{eq:g1g2g3}
g_1(\tau) & = & \left \langle ( \vec r_i(t+\tau) - \vec r_i(t) )^2 \right \rangle \nonumber\\
g_2(\tau) & = & \left \langle ( \vec r_i(t+\tau) - \vec r_{\mathrm{cm}}(t+\tau) - \vec r_i(t) + \vec r_{\mathrm{cm}}(t) )^2 \right \rangle \, , \nonumber\\
g_3(\tau) & = & \left \langle ( \vec r_{\mathrm{cm}}(t+\tau) - \vec r_{\mathrm{cm}}(t) )^2 \right \rangle \nonumber\\
\end{eqnarray}
where brackets ``$\langle ... \rangle$'' mean average over MD trajectory.
Further average over monomer position $i$ along the chain is implicitly assumed.
After complete chain relaxation, {\it i.e.} for long enough simulations, $g_1(\tau) \sim g_3(\tau) \sim \tau$ and $g_2(\tau) \simeq 2 \langle R_g^2(N) \rangle$~\cite{KremerGrestJCP1990}
where $\langle R_g^2(N) \rangle$ is the chain square gyration radius at equilibrium.
As shown in Fig.~S1 and Fig.~S2, these conditions are met for polymer solutions of chain lengths with $N=250$ and $N=500$,
while they hold less well for the largest chain size of $N=1000$, see Fig.~S3.
Nonetheless, we have decided to include data for $N=1000$ in our analysis.

\section{Results and discussion}\label{sec:Results}

Here, we will analyse and discuss our results on the statics and dynamics of entangled polymer chains in solution,
and how entanglements affect diffusion of large colloid particles.
For the sake of notation, in the rest of the paper we will denote by $L=N\sigma$ (respectively, $l=n\sigma$) the total contour length of the chain made by $N$ monomers
(respectively, the contour length of a subchain made of $n$ monomers).

We first need to provide a justification for why our polymers can effectively be considered as ``entangled''.
Consistent with previous work~\cite{RosaPLOS2008,RosaEveraersPRL2014},
we adopt here the formalism by Uchida {\it et al.}~\cite{uchida} showing
that the entanglement length of the polymer solution, $L_e$,
can be expressed as a simple function of the polymer Kuhn length, $l_K$, and the solution density, $\rho$:
$L_e / l_K = (0.06(\rho \sigma l_K^2))^{-2/5} + (0.06(\rho \sigma l_K^2))^{-2}$.
With solution densities $\rho \sigma^3 = 0.1,0.2,0.3,0.4$ and chain Kuhn length $l_K / \sigma = 10$,
corresponding entanglement lengths are given by $L_e / l_K \approx 4.00, 1.62, 1.10, 0.89$.
At the highest density then, our longest chains have a contour length $L / L_e \approx 100$.

\subsection{Chain statics}\label{sec:ChainStr}

We have computed first the mean-square spatial distance $\langle R^2(l) \rangle$ between monomers located at contour length separation $l$ along the chain.
\begin{figure}
\begin{center}
\includegraphics[width=3.1in]{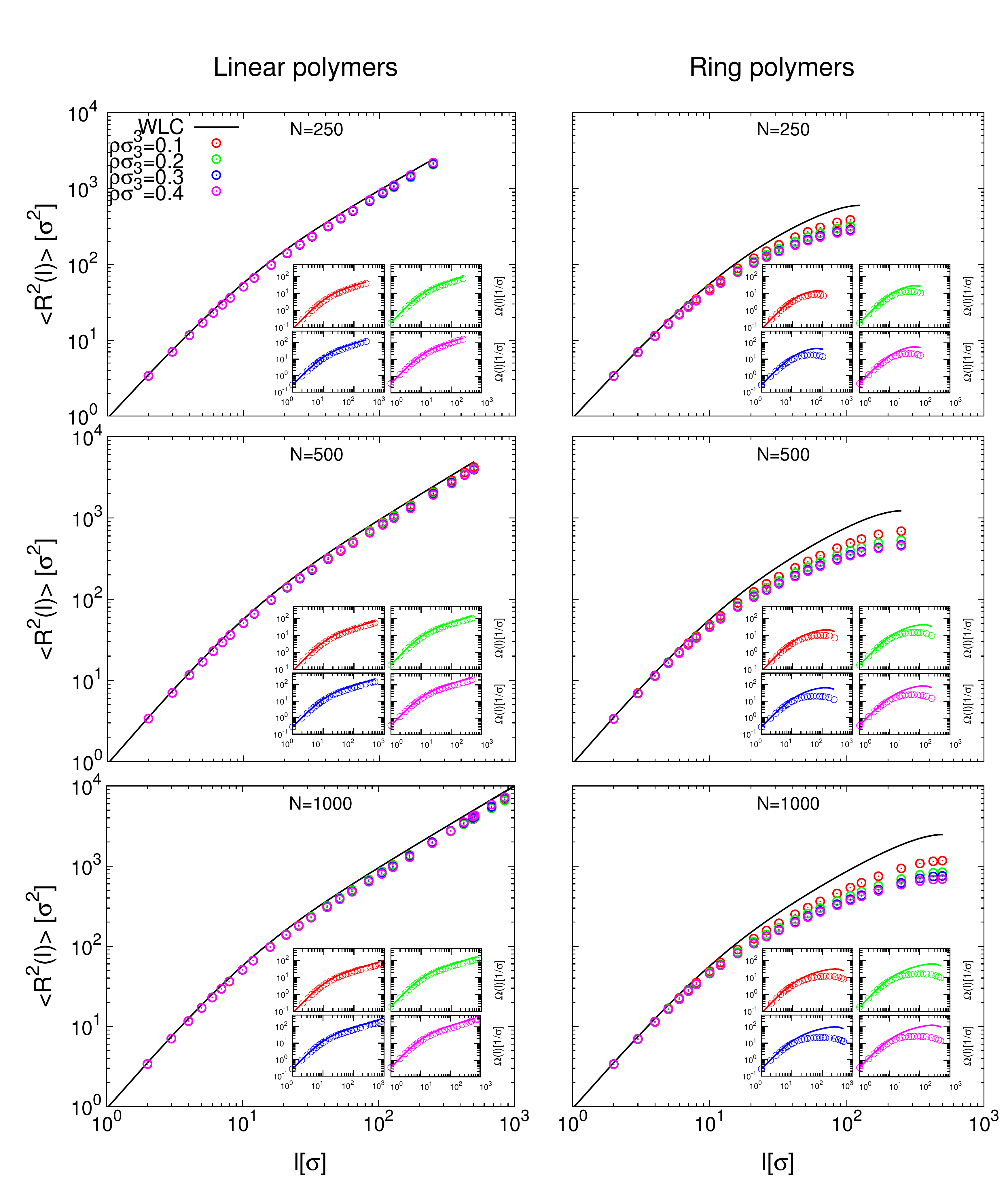}
\end{center}
\caption{
\label{fig:IntDists}
Average-square internal distances $\langle R^2(l) \rangle$ between chain monomers at contour length separation $l$ (symbols),
and corresponding theoretical predictions for worm-like chains and rings (black lines, Eqs.~\ref{eq:wlc} and~\ref{eq:wlr} respectively) with Kuhn length $l_K=10\sigma$.
Insets: corresponding overlap parameters $\Omega(l) = \rho \, \langle R^2(l) \rangle^{3/2} / l$.
}
\end{figure}
It is interesting to compare numerical results for linear and ring polymers to, respectively,
the exact worm-like chain (WLC) expression~\cite{RubinsteinColby} for semi-flexible linear polymers with Kuhn length $l_K$:
\begin{equation}\label{eq:wlc}
\langle R^2(l) \rangle^{\mathrm{WLC}} = \frac{l_K^2}{2} \left( \frac{2 \, l}{l_K} + e^{-2 \, l/l_K} - 1 \right) \, ,
\end{equation}
and the approximate formula for ideal semi-flexible rings:
\begin{equation}\label{eq:wlr}
\langle R^2(l) \rangle^{\mathrm{WLR}} = \left( \frac{1}{\langle R^2(l) \rangle^{\mathrm{WLC}}} + \frac{1}{\langle R^2(L-l) \rangle^{\mathrm{WLC}}} \right)^{-1} \, 
\end{equation}
which gives an accurate description provided $L \gg l_K$~\cite{NoteOnWLREq}.
Numerical results and analytical expressions for $\langle R^2(l) \rangle$ are summarised in Fig.~\ref{fig:IntDists} as symbols and black solid lines, respectively.
We remark the striking difference between linear chains and rings:
as expected based on the theoretical scenario~\cite{DoiEdwards} predicting the screening of excluded volume interactions,
numerical results for $\langle R^2(l) \rangle$ of linear chains (Fig.~\ref{fig:IntDists}, left panels) show little or no dependence on density and agree well with the WLC prediction, Eq.~\ref{eq:wlc}.
Conversely, in solutions of ring polymers where screening is absent~\cite{mullerPRE1996,mullerPRE2000,Vettorel2009,DeguchiJCP2009,Halverson2011_1,HalversonPRL2012,RosaEveraersPRL2014}
numerical predictions for $\langle R^2(l) \rangle$ markedly deviate from Eq.~\ref{eq:wlr} (right panels of Fig.~\ref{fig:IntDists}).
In particular, rings show the tendency of becoming more and more compact as density increases.
Alternatively~\cite{RosaEveraersPRL2014}, the same data can be recast in terms of the so-called overlap parameter
$\Omega(l) \equiv \rho \, \langle R^2(l) \rangle^{3/2} / l$
which gives the total number of sub-chains of contour length $l$ contained inside the corresponding occupied volume.
Results are shown as insets in Fig.~\ref{fig:IntDists}, highlighting the important difference between chains and rings:
in fact, for the latter $\Omega(l)$ tends to plateau at increasing chain size ({\it i.e.} $\langle R^2(l) \rangle \sim l^{2\nu}$ with critical exponent~\cite{RubinsteinColby} $\nu=1/3$), the value of the plateau slightly increasing from $\approx 12$ for $\rho=0.1$ to $\approx 28$ for $\rho=0.4$.

\begin{figure}
\includegraphics[width=3.2in]{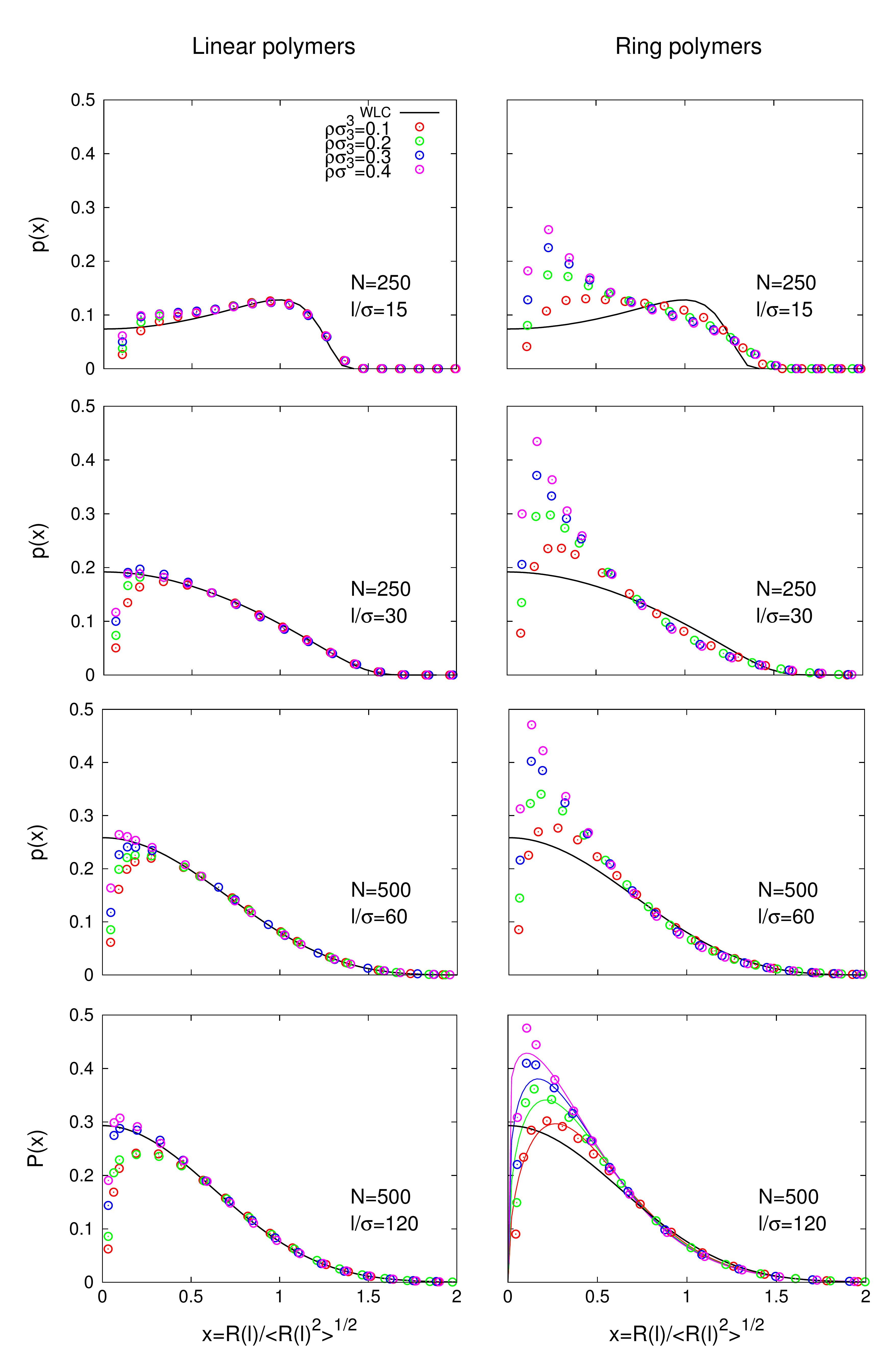}
\caption{
\label{fig:End2EndPDF}
Distribution functions of end-to-end distances, $R(l)$ at fixed contour length separations, $l = 15, 30, 60, 120$.
Black lines correspond to the semi-empirical WLC formula, Eq.~\ref{eq:BeckerWLC}.
Solid coloured lines in the bottom right panel corresponds to the RdC function, Eq.~\ref{eq:RdC}, with fit parameters given in Eq.~\ref{eq:RdCfit}.
}
\end{figure}

Next, we have considered the full distribution functions $p(R | l)$ of end-to-end distances $R=R(l)$ as a function of $l=15,30,60,120$ and density $\rho$, see Fig.~\ref{fig:End2EndPDF}.
Both linear and ring polymers show the expected shift from the non-universal short chain behaviour ($l=15$) where fiber stiffness plays the dominant role to the universal, entropy-governed long chain behaviour.
Again, there are important noticeable differences between linear and ring polymers:
For linear chains, the progressive screening of excluded volume effects~\cite{DoiEdwards} at increasing $\rho$ is well exemplified by the fact that $p(R|l)$ super-imposes on the semi-empirical formula of ideal worm-like chains given in~\cite{BeckerRosaEveraers} (left panels in Fig.~\ref{fig:End2EndPDF}, black lines):
\begin{eqnarray}\label{eq:BeckerWLC}
p(R | l)^{\mathrm{WLC}}
& = & J(l) \left( \frac{1 - c R^2}{1-R^2} \right)^{5/2} e^{\frac{\sum_{i=-1}^0 \sum_{j=1}^3 c_{ij} \left(\frac{l_K}{2 l}\right)^i R^{2j}}{1-R^2}} \nonumber\\
& & \times \, e^{-\frac{d \, \frac{l_K}{l} \, ab(1+b)R^2}{1-b^2 R^2}} \, I_0 \left( -\frac{d \, \frac{l_K}{l} \, a(1+b)R}{1-b^2 R^2} \right) \nonumber\\
\end{eqnarray}
with numerical constants $a=7.027$, $b=0.473$, and $c_{ij} = \left( \begin{array}{ccc} -3/4 & 23/64 & -7/64 \\ -1/2 & 17/16 & -9/16 \end{array} \right)$,
$I_0$ the modified Bessel function of the first kind and
\begin{eqnarray}\label{eq:BeckerWLC_details}
1-c & = & \left(1 + \left( 0.734 \left( \frac{l_K}{l} \right)^{-0.95} \right)^{-5} \right)^{-1/5} \nonumber\\
\nonumber\\
\nonumber\\
1-d & = & \left\{ \begin{array}{ll} 0 , & \frac{l_K}{l} < \frac{1}{4} \\ \\ \frac{1}{\frac{0.354}{l_K/l-0.222} + 3.719\left(\frac{l_K}{l}-0.222\right)^{0.783}} , & \mbox{otherwise} \end{array}\right. \nonumber\\
\nonumber\\
J(l) & = & \left\{ \begin{array}{ll} 28.01 \left(\frac{l_K}{l}\right)^2 e^{0.492 \frac{l}{l_K} - a \frac{l_K}{l}} , & \frac{l_K}{l} > \frac{1}{4} \\ \\ \left( \frac{3 \, l}{2 \pi \, l_K}\right)^{3/2}\left( 1-\frac{5 \, l_K}{8 l}\right) , & \frac{l_K}{l} \leq \frac{1}{4} \end{array} \right.\nonumber\\
\end{eqnarray}
In particular, for large $l$'s $p(R | l)$ becomes {\it nearly} Gaussian (Fig.~\ref{fig:End2EndPDF}, last panel on the left).
In striking contrast, the large-$l$ behaviour of $p(R | l)$ for ring polymers is markedly non-Gaussian.
In particular, $p(R | l=120)$ for well equilibrated rings with $N=500$ are well described (solid lines, last panel on the right of Fig.~\ref{fig:End2EndPDF}) by the classical Redner-des Cloizeaux (RdC) function~\cite{Redner1980,DesCloizBook,EveraersJPA1995,RosaGrosbergEveraers2015}:
\begin{eqnarray}\label{eq:RdC}
p(R | l)^{\mathrm{RdC}} &=& \frac1{\langle R^2(l) \rangle^{3/2}} \ q\left(\frac{R(l)}{ \sqrt{\langle R^2(l) \rangle}}\right) \nonumber\\
q(\vec x) &=&  C \, x^\theta\  \exp \left( -K x^t \right) \, ,
\end{eqnarray}
where the two constants $C$ and $K$ are determined by the conditions
(1) that the distribution is normalized ($\int q(x) \, 4 \pi x^2 \, dx \equiv 1 $) and
(2) that the second moment was chosen as the scaling length ($\int x^2 q(x)  \, 4 \pi x^2 \, dx \equiv 1 $):
$C = t \, \frac{\Gamma(\frac{5}{2})\Gamma^{\frac{3+\theta}2}(\frac{5+\theta}t)}{3\,\pi^{3/2}\,\Gamma^{\frac{5+\theta}2}(\frac{3+\theta}t)}$ and
$K^2 =  \frac{\Gamma(\frac{5+\theta}t)}{\Gamma(\frac{3+\theta}t)}$.
Fit of data with $l=120$ to the two-parameter RdC function gives the following estimates for $\theta$ and $t$~\cite{FitDetailsNote}:
\begin{equation}\label{eq:RdCfit}
\begin{array}{ccc}
\rho\sigma^3=0.1: & \theta=0.5 \pm 0.1 , & t=1.6 \pm 0.1 \\
\rho\sigma^3=0.2: & \theta=0.4 \pm 0.2, & t=1.5 \pm 0.2\\
\rho\sigma^3=0.3: & \theta=0.3 \pm 0.2 , & t=1.5 \pm 0.2 \\
\rho\sigma^3=0.4: & \theta=0.1 \pm 0.2 , & t=1.5 \pm 0.2
\end{array}
\end{equation}
Interestingly, $t$ appears compatible with $3/2$, a result which is consistent with the Fisher-Pincus~\cite{FisherSAWShape1966,PincusBlob1976} relationship $t = 1 / (1-\nu)$ with $\nu=1/3$.
On the other hand, the excluded-volume exponent~\cite{DesCloizBook} $\theta$ tends to become small as density increases,
finally suggesting for the asymptotic high-density limit the simple and elegant stretched-exponential form
$q(\vec x) \sim \exp(-K x^{3/2})$ with $K = \sqrt{ \frac{\Gamma(10/3)}{\Gamma(2)} } \approx 1.667$.

\begin{figure}
\includegraphics[width=3.2in]{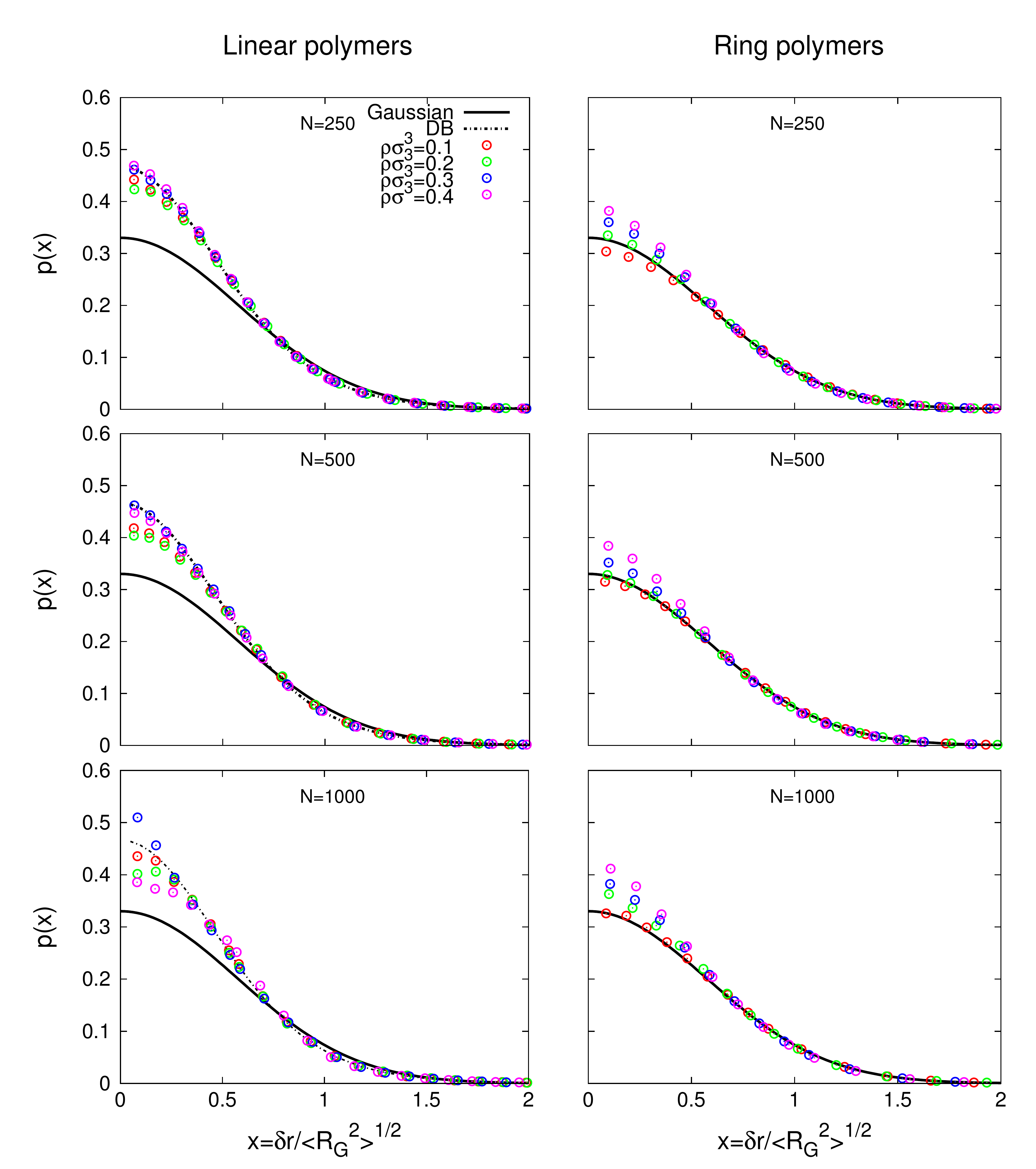}
\caption{
\label{fig:CoMPDF}
Probability distribution function of monomer spatial distances from the chain center of mass, $\delta r$ (symbols),
in comparison to the Gaussian distribution function (black solid lines), Eq.~\ref{eq:Gaussian},
and the analytical distribution function by Debye and Bueche (black dashed lines), Eq.~\ref{eq:DebyeBueche}.
}
\end{figure}

We continue our analysis on chains structure by studying (Fig.~\ref{fig:CoMPDF}) the probability distribution functions $p(\vec{\delta r})$ of monomer spatial distances from the chain center of mass
$\vec{\delta r} \equiv {\vec r} - {\vec r}_{\mathrm{cm}}$,
whose second moment corresponds to the mean-square gyration radius $\langle R_g^2(L) \rangle$.
Again, linear chains and rings have different behaviours.
Consistent with the generic result for worm-like chains~\cite{DebyeBuecheJCP1952},
linear chains deviate from the Gaussian function (black solid lines):
\begin{equation}\label{eq:Gaussian}
p(\vec{\delta r})^G = \left( \frac{3}{2 \pi \langle R_g^2(L) \rangle} \right)^{3/2} \exp\left( -\frac{3 (\vec{\delta r})^2}{2 \langle R_g^2(L) \rangle}\right) \, ,
\end{equation}
and agree well with the exact analytical prediction by Debye and Bueche~\cite{DebyeBuecheJCP1952} (black dashed lines):
\begin{equation}\label{eq:DebyeBueche}
p(\vec{\delta r})^{\mathrm{DB}} = \frac{1}{L} \int_0^L dl \left( \frac{3}{2 \pi \langle (\vec{\delta r_l})^2 \rangle} \right)^{3/2} \exp\left( -\frac{3 (\vec{\delta r_l})^2}{2 \langle (\vec{\delta r_l})^2 \rangle}\right) \, ,
\end{equation}
where $\langle (\vec{\delta r_l})^2 \rangle = 2 \langle R_g^2(L) \rangle \left( 1 - 3 \frac{l}{L} \left( 1 - \frac{l}{L} \right) \right)$, with $\langle R_g^2(L) \rangle = l_K L / 6$.
Interestingly, distribution functions for ring polymers are closer to the Gaussian distribution, although with some noticeable deviations.

\begin{figure}
\includegraphics[width=2.4in,angle=90]{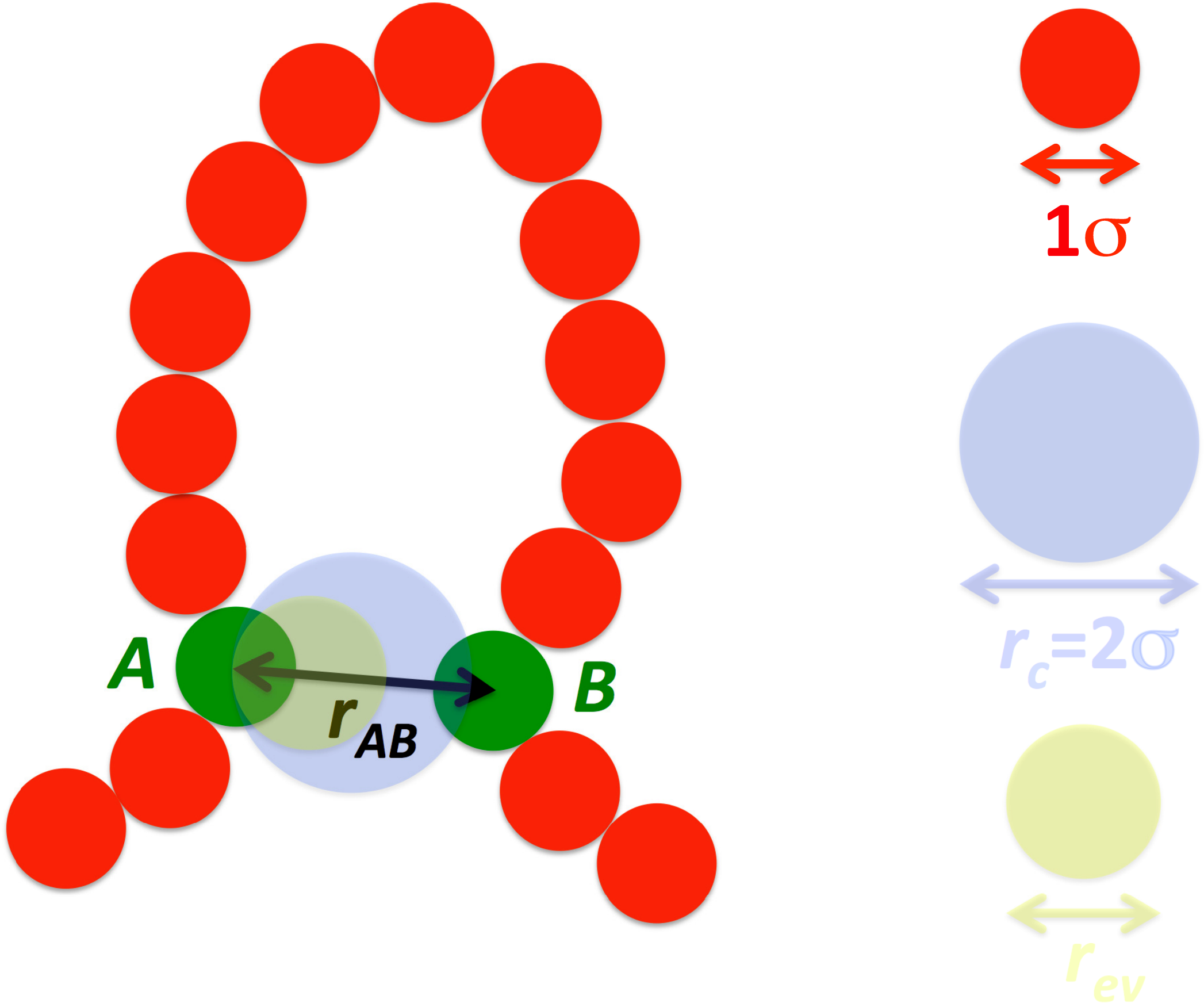}
\caption{
\label{fig:EVExplanatoryFig}
Two monomers $A$ and $B$ (in green) of diameter $\sigma$ along a polymer chain (in red) are said to be in contact if their centre-to-centre spatial distance $r_{\mathrm{AB}}$ is smaller than $r_c=2\sigma$.
Because of excluded volume effects, two monomers can not be closer than a certain distance $r_{\mathrm{ev}}$, which is a decreasing function of solution density,
see text for details.
}
\end{figure}

\begin{figure}
\includegraphics[width=3.2in]{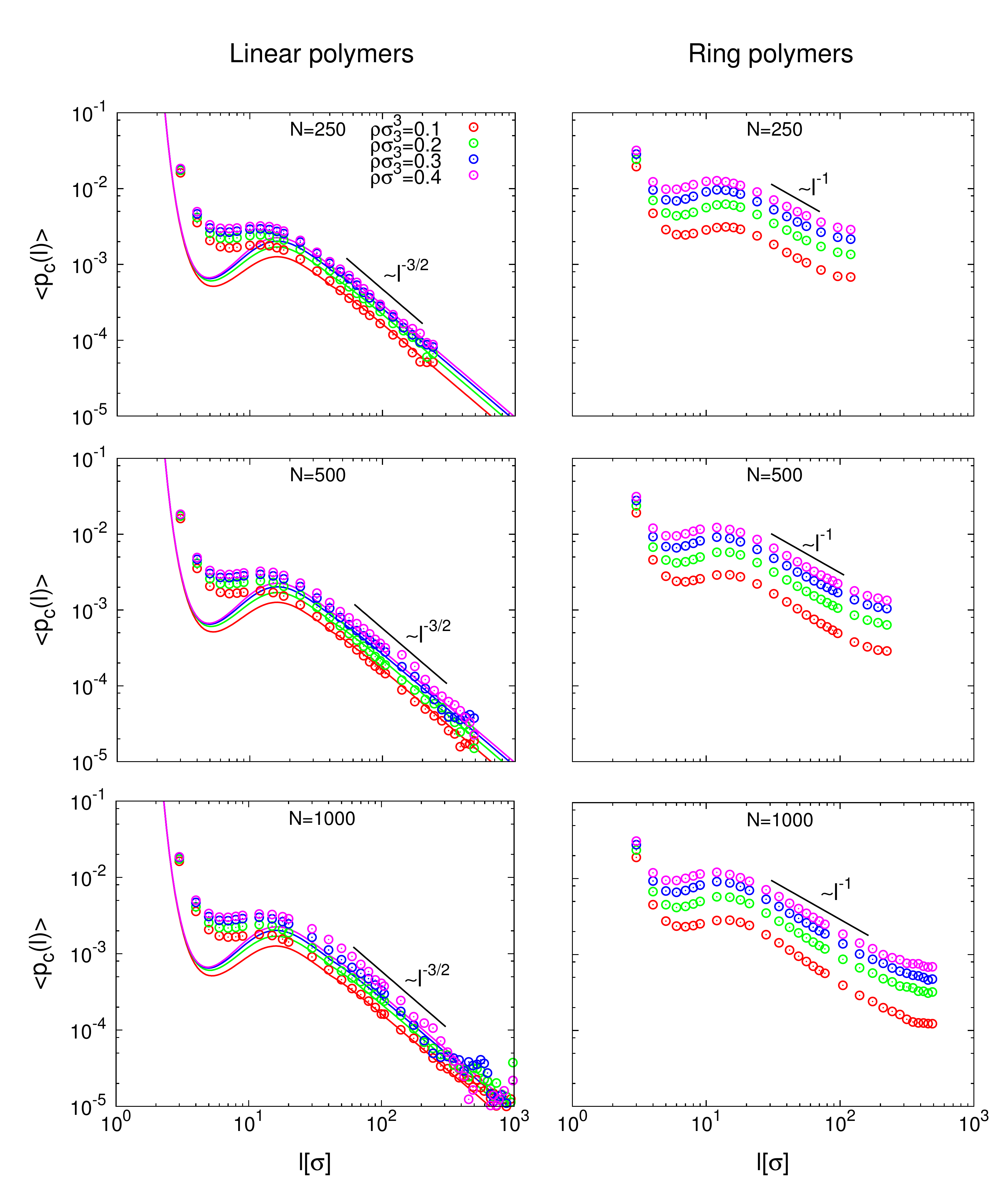}
\caption{
\label{fig:ContactFreqs}
Average-square contact frequencies $\langle p_c(l) \rangle$ between monomers at contour length separation, $l$.
Solid lines in left panels correspond to numerical integration of Eq.~\ref{eq:ContactFreqsDef} with $p(R | l)$ given by the semi-empirical WLC formula, Eq.~\ref{eq:BeckerWLC}.
}
\end{figure}

\begin{figure}
\includegraphics[width=3.2in]{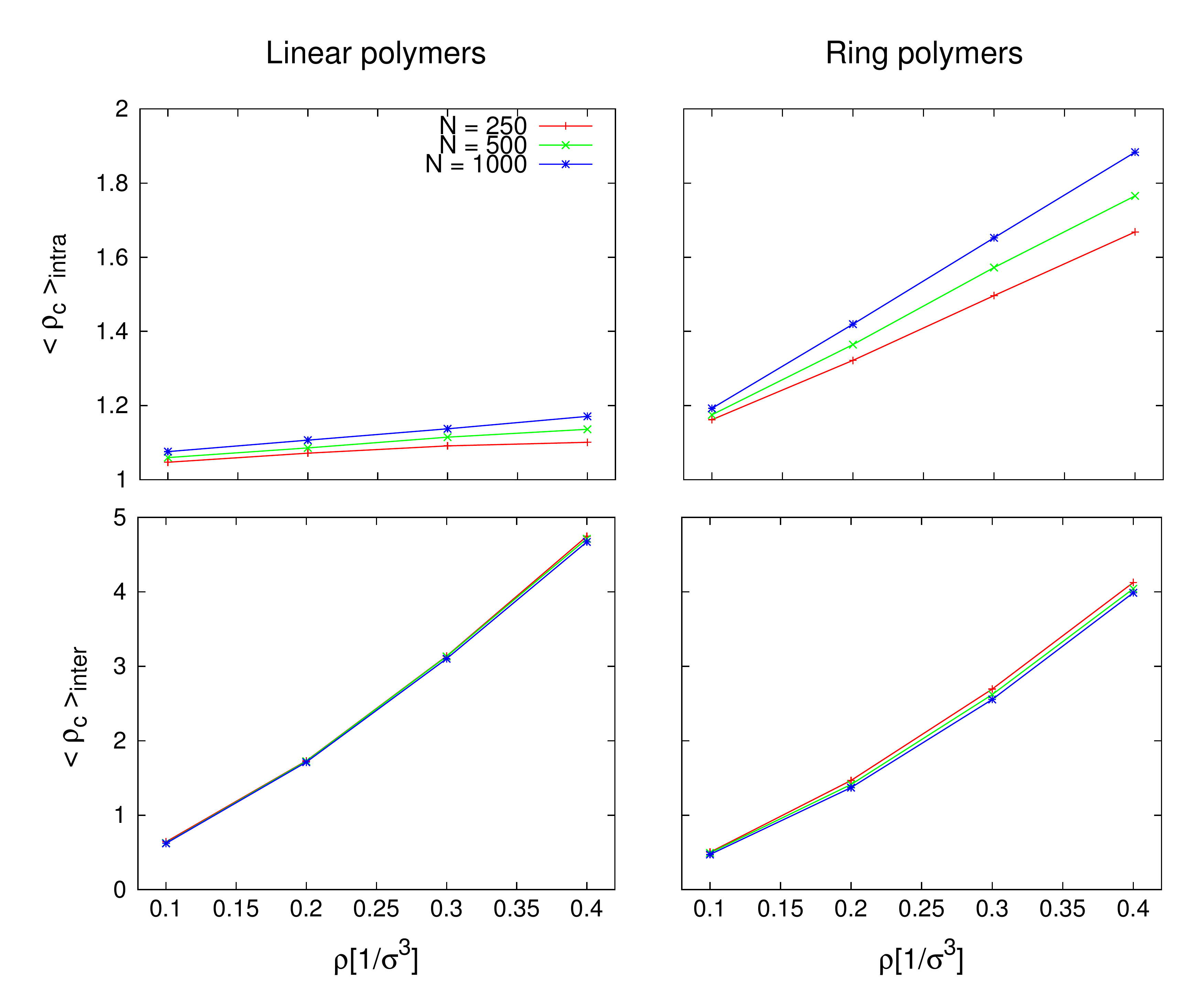}
\caption{
\label{fig:IntraVsInterContacs}
Average number of contacts per chain monomer:
separate contributions arising from contacts between monomers inside the same chain, $\langle \rho_c \rangle_{\mathrm{intra}}$ (top panels),
and
from contacts between monomers belonging to different chains, $\langle \rho_c \rangle_{\mathrm{inter}}$ (bottom panels).
}
\end{figure}

We have studied next the effect of topological constraints on the average contact frequencies, $\langle p_c(l) \rangle$, between chain monomers at contour length separation $l$ defined as:
\begin{equation}\label{eq:ContactFreqsDef}
\langle p_c(l) \rangle \equiv \frac{\int_{r_{\mathrm{ev}}}^{r_c} \, p(R | l) \, 4\pi R^2 dR}{\int_{0}^{l} \, p(R | l) \, 4\pi R^2 dR} \, ,
\end{equation}
where $r_{\mathrm{ev}}$ is the distance of closest approach due to intra-monomer excluded volume effects and $r_c = 2\sigma$ is the chosen contact cutoff distance (Fig.~\ref{fig:EVExplanatoryFig}).
For both linear and ring polymers $\langle p_c(l) \rangle$ increases as a function of density, Fig.~\ref{fig:ContactFreqs}.
For linear chains, this is arguably due to the progressive screening of excluded volume effects.
In particular, by using Eq.~\ref{eq:ContactFreqsDef} with the WLC expression Eq.~\ref{eq:wlc} the long-$l$ behaviour of $\langle p_c(l) \rangle$ can be well reproduced (solid lines in left panels of Fig.~\ref{fig:ContactFreqs}) by the following values for $r_{\mathrm{ev}}$:
$r_{\mathrm{ev}} / \sigma = 1.6, 1.4, 1.2, 1.0$ for, respectively, $\rho \sigma^3 = 0.1, 0.2, 0.3, 0.4$.
Then, $\langle p_c(l) \rangle \sim l^{-3/2}$ at all densitites.
Conversely, the observed tendency of $\langle p_c(l) \rangle$ to shift towards higher values is the consequence of rings becoming more and more compact as density increases (Fig.~\ref{fig:End2EndPDF}, right panels).
Furthermore, now the observed scaling law is different, $\langle p_c(l) \rangle \sim l^{-1}$, and in particular compatible with the predictions of crumpled globules~\cite{hic,Halverson2011_1,RosaEveraersPRL2014}.
This analysis is complemented by considering separately the two contributions to the average number of contacts of each monomer of the chain:
the first arising from contacts between monomers inside the same chain, $\langle \rho_c \rangle_{\mathrm{intra}}$,
and
the second arising from contacts between monomers belonging to different chains, $\langle \rho_c \rangle_{\mathrm{inter}}$.
The results are shown in Fig.~\ref{fig:IntraVsInterContacs}.
As expected, $\langle \rho_c \rangle_{\mathrm{intra}}$ for linear chains, show almost no variation with solution density or chain length, in agreement with the picture that chains remain nearly ideal.
On the other hand, $\langle \rho_c \rangle_{\mathrm{intra}}$ for rings increases significantly with density.
This effect can not be ascribed to local contacts along the chain (otherwise we should have seen a similar effect for linear chains, too),
while it can be easily understood in terms of the crumpling of the rings which constrains distal monomers along the chain to move close in space.
The second contribution, $\langle \rho_c \rangle_{\mathrm{inter}}$, increases in a similar manner for both linear and ring polymers,
demonstrating in particular that crumpling does not prevent a single ring to maintain substantial interactions with its neighbours.

\subsection{Chain and colloid dynamics}\label{sec:Dynamics}

Chain dynamics can be appropriately characterized by considering the following three quantities~\cite{DoiEdwards,KremerGrestJCP1990,Halverson2011_2}:
the mean-square displacement of single monomers in the absolute frame ($g_1(\tau)$) and in the frame of the chain center of mass ($g_2(\tau)$),
and the mean-square displacement of the chain center of mass ($g_3(\tau)$) at lag-time $\tau$ (Eqs.~\ref{eq:g1g2g3} for definitions),
see Figs.~S1-S3.
In the long-time limit, $g_1(\tau) \sim g_3(\tau) \sim \tau$
and $g_2(\tau) \sim 2 \langle R_g^2(L) \rangle$~\cite{KremerGrestJCP1990} where $\langle R_g^2(L) \rangle$ is the mean-square gyration radius of the chains.
We notice that, for all cases considered ring polymers relax faster than linear ones.
Particularly striking is the example of $g_2(\tau)$ for $N=1000$ linear chains at $\rho\sigma^3 = 0.1$ (Fig.~S3)
showing that present simulations are still too short to guarantee complete relaxation of linear chains, but long enough to equilibrate their circular counterparts.
These results are in agreement with theoretical arguments by Rubinstein and coworkers~\cite{ORD_PRL1994} related to the different ways entanglements affect relaxation of linear chains and rings.
Linear chains obey the classical mechanism of ``reptation''~\cite{deGennes71,DoiEdwards}:
hindered by topological constraints they reduce to slide past each others in a ``snake-like'' fashion, the total relaxation time being $\sim L^3$.
Pointing out on the analogy between entangled ring polymers and branched trees,
Rubinstein and coworkers suggested~\cite{ORD_PRL1994} that ring polymers relax instead owing to mass flowing along the ring contour length,
a mechanism leading to the faster relaxation time $\sim L^{2.5}$.

\begin{figure}
\includegraphics[width=3.2in]{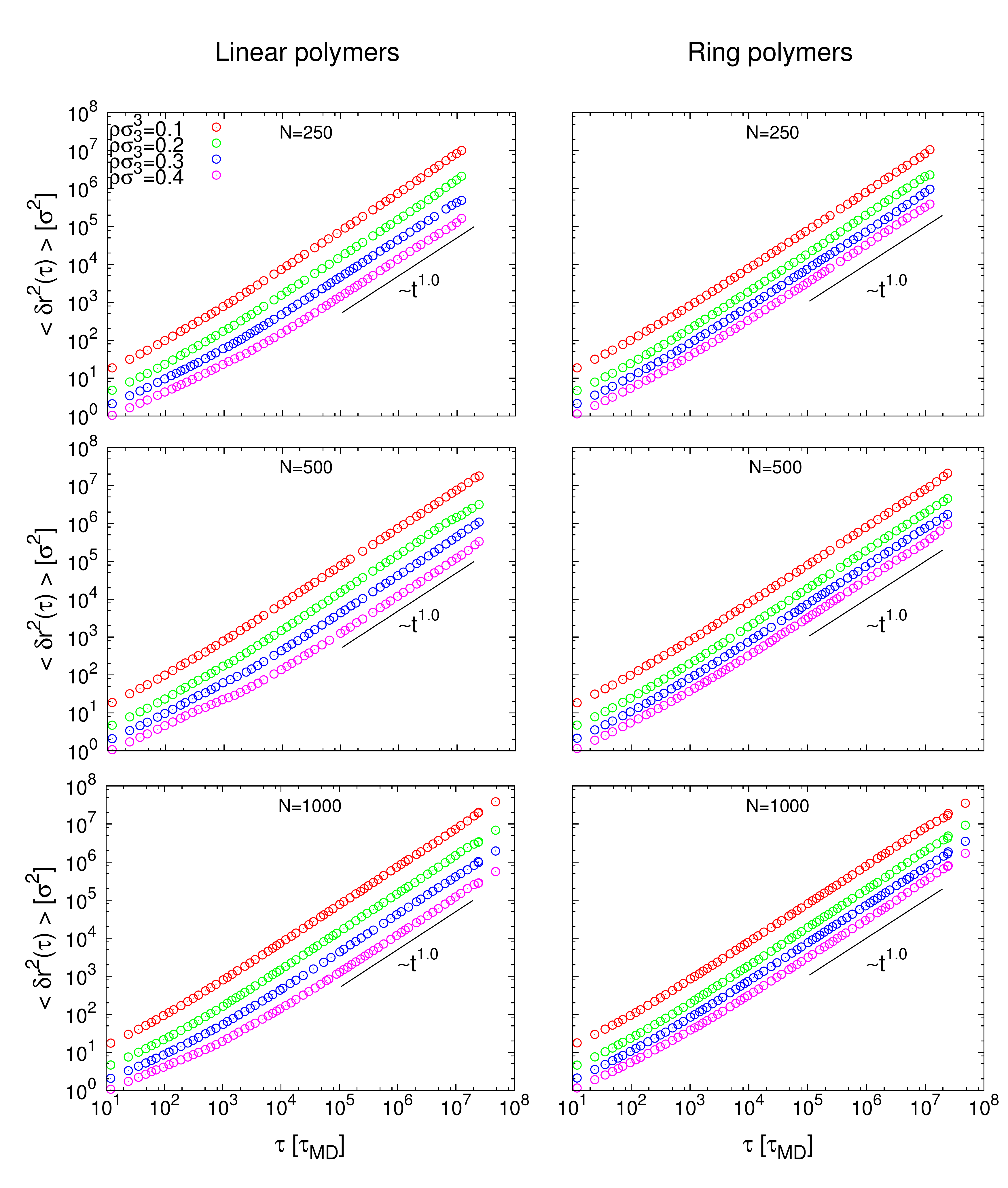}
\caption{
\label{fig:ColloidMSD}
Mean-square displacement, $\langle \delta r^2(\tau) \rangle$, of dispersed colloid particles of diameter $d=5\sigma$.
}
\end{figure}

\begin{table*}
\begin{tabular}{|c|c|c|c|c|c|c|}
\hline
\multicolumn{7}{|c|}{Linear polymers}\\
\hline
& \multicolumn{2}{c|}{$N=250$} & \multicolumn{2}{c|}{$N=500$} & \multicolumn{2}{c|}{$N=1000$}\\
\hline
{$\rho \sigma^3$} &
$D_{\infty} \, [\frac{\sigma^2}{\tau_{MD}}]$ & $\eta_{\infty} \, [\frac{k_B T \, \tau_{MD}}{\sigma^3}]$ &
$D_{\infty} \, [\frac{\sigma^2}{\tau_{MD}}]$ & $\eta_{\infty} \, [\frac{k_B T \, \tau_{MD}}{\sigma^3}]$ &
$D_{\infty} \, [\frac{\sigma^2}{\tau_{MD}}]$ & $\eta_{\infty} \, [\frac{k_B T \, \tau_{MD}}{\sigma^3}]$ \\
\hline
$0.1$ & $(1.41 \pm 0.01) \times 10^{-1}$ & $0.188 \pm 0.001$ & $(1.26 \pm 0.01) \times 10^{-1}$ & $0.210 \pm 0.002$ & $(1.34 \pm 0.01) \times 10^{-1}$ & $0.198 \pm 0.002$ \\
$0.2$ & $(2.88 \pm 0.04) \times 10^{-2}$ & $0.92 \pm 0.01$ & $(2.34 \pm 0.03) \times 10^{-2}$ & $1.13 \pm 0.02$ & $(2.39 \pm 0.03) \times 10^{-2}$ & $1.11 \pm 0.01$ \\
$0.3$ & $(7.1 \pm 0.2) \times 10^{-3}$ & $3.7 \pm 0.1$ & $(7.5 \pm 0.1) \times 10^{-3}$ & $3.53 \pm 0.02$ & $(6.8 \pm 0.1) \times 10^{-3}$ & $3.90 \pm 0.04$ \\
$0.4$ & $(2.19 \pm 0.04) \times 10^{-3}$ & $12.1 \pm 0.2$ & $(2.2 \pm 0.1) \times 10^{-3}$ & $12.0 \pm 0.3$ & $(1.97 \pm 0.03) \times 10^{-3}$ & $13.4 \pm 0.2$ \\
\hline
\multicolumn{7}{c}{}\\
\hline
\multicolumn{7}{|c|}{Ring polymers}\\
\hline
& \multicolumn{2}{c|}{$N=250$} & \multicolumn{2}{c|}{$N=500$} & \multicolumn{2}{c|}{$N=1000$}\\
\hline
{$\rho \sigma^3$} &
$D_{\infty} \, [\frac{\sigma^2}{\tau_{MD}}]$ & $\eta_{\infty} \, [\frac{k_B T \, \tau_{MD}}{\sigma^3}]$ &
$D_{\infty} \, [\frac{\sigma^2}{\tau_{MD}}]$ & $\eta_{\infty} \, [\frac{k_B T \, \tau_{MD}}{\sigma^3}]$ &
$D_{\infty} \, [\frac{\sigma^2}{\tau_{MD}}]$ & $\eta_{\infty} \, [\frac{k_B T \, \tau_{MD}}{\sigma^3}]$ \\
\hline
$0.1$ & $(1.43 \pm 0.02) \times 10^{-1}$ & $0.186 \pm 0.003$ & $(1.41 \pm 0.04) \times 10^{-1}$ & $0.19 \pm 0.01$ & $(1.23 \pm 0.02) \times 10^{-1}$ & $0.216 \pm 0.002$ \\
$0.2$ & $(3.4 \pm 0.1) \times 10^{-2}$ & $0.79 \pm 0.02$ & $(3.12 \pm 0.02) \times 10^{-2}$ & $0.85 \pm 0.01$ & $(3.2 \pm 0.1) \times 10^{-2}$ & $0.83 \pm 0.01$ \\
$0.3$ & $(1.32 \pm 0.01) \times 10^{-2}$ & $2.00 \pm 0.02$ & $(1.21 \pm 0.01) \times 10^{-2}$ & $2.20 \pm 0.02$ & $(1.21 \pm 0.02) \times 10^{-2}$ & $2.2 \pm 0.03$ \\
$0.4$ & $(5.46 \pm 0.02) \times 10^{-3}$ & $4.86 \pm 0.02$ & $(5.7 \pm 0.1) \times 10^{-3} $ & $4.6 \pm 0.1$ & $(5.7 \pm 0.1) \times 10^{-3}$ & $4.7 \pm 0.1$ \\
\hline
\end{tabular}
\caption{
\label{tab:DynObs}
Dynamical properties of semi-flexible linear chains and ring polymers in solution of density $\rho$:
$D_{\infty}$: terminal diffusion coefficient of dispersed colloid particle of diameter $d=5\sigma$;
$\eta_{\infty}$: corresponding terminal viscosity.
}
\end{table*}

We now discuss the micro-rheological properties of polymer solutions by studying the diffusive motion of hard-sphere colloid particles~\cite{TsengWirtz2004,WirtzReview}
of diameter $d=5\sigma$.
This size was specifically chosen because, being at the crossover between the solution mesh size of $\approx 8 \sigma$ for $\rho\sigma^3 = 0.1$ and $\approx 4 \sigma$ for $\rho\sigma^3 = 0.4$~\cite{NoteOnMeshSize},
it ought to be relevant to characterise how entanglements affect the local viscosity of solutions of linear and ring polymers.
Fig.~\ref{fig:ColloidMSD} shows our final results for the mean-square displacement $\langle \delta r^2(\tau) \rangle \equiv \langle ({\vec r}_{c}(t+\tau) - {\vec r}_{c}(t))^2 \rangle$
of colloid position ${\vec r}_{c}(t)$ at lag-time $\tau$.
At the lowest density, colloids diffusion does not depend on the chain topology.
Conversely, as the highest density is attained colloids diffuse manifestly faster in solution of ring polymers.
In either cases, $\langle \delta r^2(\tau) \rangle \sim \tau$ at long-times.
We characterise then this density-dependent long-time regime
by defining the corresponding diffusion coefficients
$D_{\infty}(\rho) \equiv \lim_{\tau=\infty} \frac{\langle \delta r^2(\tau) \rangle}{6 \tau}$
and viscosities $\eta_{\infty}$ as given by the classical Stokes-Einstein relationship~\cite{DoiEdwards}
$\eta_{\infty}(\rho) = \frac{k_B T}{2 \pi \, (d+\sigma) \, D_{\infty}(\rho)}$~\cite{NoteOnBoundaryConditions}.
Values for $D_{\infty}(\rho)$ obtained by best fits of the data and corresponding $\eta_{\infty}(\rho)$ (which are $N$-independent!) are reported in Table~\ref{tab:DynObs}.
We thus see that solutions of ring polymers are up to $\approx 3$ times less viscous than their corresponding linear counterparts.

\section{Conclusion}\label{sec:Concls}
In this article, we have presented the results of Molecular Dynamics computer simulations for the characterization of the statics and dynamics
of semi-flexible linear and ring polymers in semi-dilute solutions.
Chains of different sizes and at different solution densities have been considered.

In agreement with the well known picture invoking screening~\cite{DoiEdwards} of excluded volume effects,
we confirm that linear chains behave as quasi-ideal at all considered densities (Fig.~\ref{fig:IntDists}, left).
Conversely (Fig.~\ref{fig:IntDists}, right), ring polymers at same physical conditions tend to become increasingly more compact.
These results prompted us to consider the full chain statistics given by the distribution function $p(R|l)$ of spatial distances between the ends of subchain of linear size $l$ (Fig.~\ref{fig:End2EndPDF}):
at high densities screening effects in linear chains extend down to small scales and $p(R|l)$ is well described by the worm-like chain statistics. In particular, at large $l$'s chain statistics is almost Gaussian.
On the other hand, chain compaction in rings induces deviations from the ideal statistics at all $l$'s.
Interestingly, by describing the large-$l$ behavior of $p(R|l)$ by the classical Redner-des Cloizeaux~\cite{Redner1980,DesCloizBook} statistics we suggest that $p(R|l)$ ought to obey the universal stretched exponential form $\sim \exp \left[ -1.667 \, \left( \frac{R(l)}{\langle R^2(l) \rangle^{1/2}} \right)^{3/2} \right]$, which satisfies the Fisher-Pincus relationship~\cite{FisherSAWShape1966,PincusBlob1976}.
We finalised the description of chain statistics by measuring the frequencies of monomer-monomer interactions inside the same chain (Fig.~\ref{fig:ContactFreqs}) and between different chains (Fig.~\ref{fig:IntraVsInterContacs}).
In particular, we explain the observed increasing of contact frequencies with the solution density (at fixed chain length $l$) by taking into account the progressive screening of excluded volume effects (for linear polymers)
and chain compaction (for ring polymers).

Finally, we have investigated chain relaxation at equilibrium (Figs.~S1-S3), confirming the general tendency~\cite{ORD_PRL1994,Halverson2011_2} of rings to relax faster than linear chains of the same size.
At the same time, we provide previously unreported evidence that dispersed colloid particles of linear size exceeding the nominal mesh size of the solution show the tendency to diffuse more rapidly in ring polymers solutions than in linear polymers ones.
Intriguingly, it was recently pointed out~\cite{RosaPLOS2008,Grosberg_PolSciC_2012} that the experimentally observed chromosome organization in eukaryotes should resemble a solution of ring polymers.
In this respect then, enzymes and other functional macromolecular complexes which need to run and bind to specific
target sequences along the genome might have greatly benefitted from the enhanced mobility stemming from the peculiar organization of the genome~\cite{maeshimaCellRep2012}.

{\it Acknowledgements} --
The authors are indebted to G. D'Adamo for useful discussions and suggestions.
They acknowledge computational resources from SISSA HPC facilities.
AR acknowledges grant PRIN 2010HXAW77 (Ministry of Education, Italy).


\clearpage

\setcounter{section}{0}
\setcounter{figure}{0}
\setcounter{table}{0}
\setcounter{equation}{0}

\renewcommand{\figurename}{Fig. S}
\renewcommand{\tablename}{Table S}

{\large \bf Supplemental Material}

\tableofcontents

\begin{figure*}
$$
\begin{array}{c}
\includegraphics[width=5.5in]{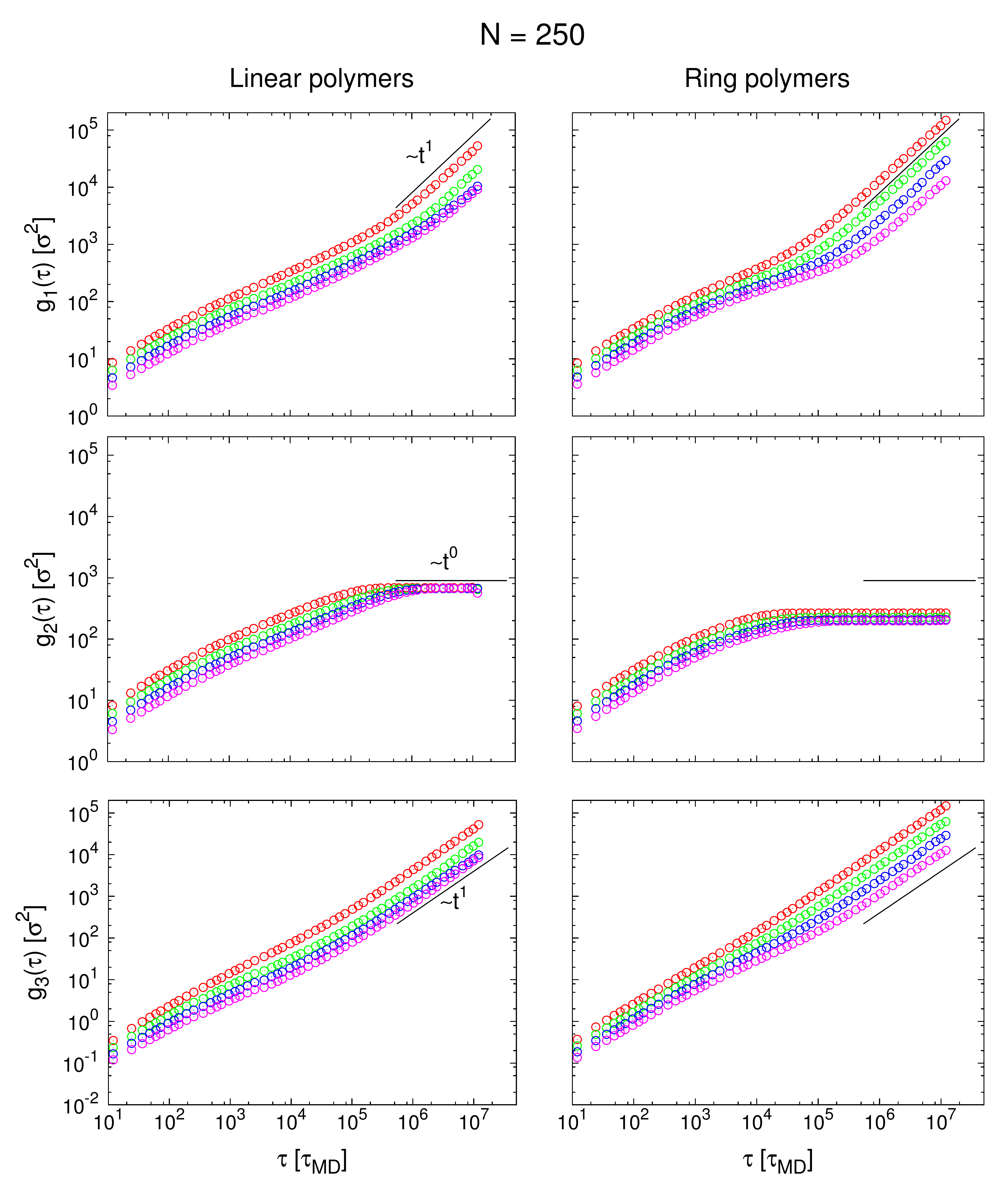}
\end{array}
$$
\caption{
\label{fig:monomerMSD_N250}
Dynamics of chains made of $N=250$ monomers.
Time behaviours for
the monomer mean-square displacement in the absolute frame ($g_1(\tau)$),
the monomer mean-square displacement in the frame of the chain centre of mass ($g_2(\tau)$),
the mean-square displacement of the chain centre of mass ($g_3(\tau)$),
see Eqs.~8 in the main text for definitions.
The reported power-laws correspond to the expected [M. Doi and S. F. Edwards, {\it The Theory of Polymer Dynamics}, Oxford Univ. Press, New York (1986)] long time behaviour after complete chain relaxation.
}
\end{figure*}

\begin{figure*}
$$
\begin{array}{c}
\includegraphics[width=5.5in]{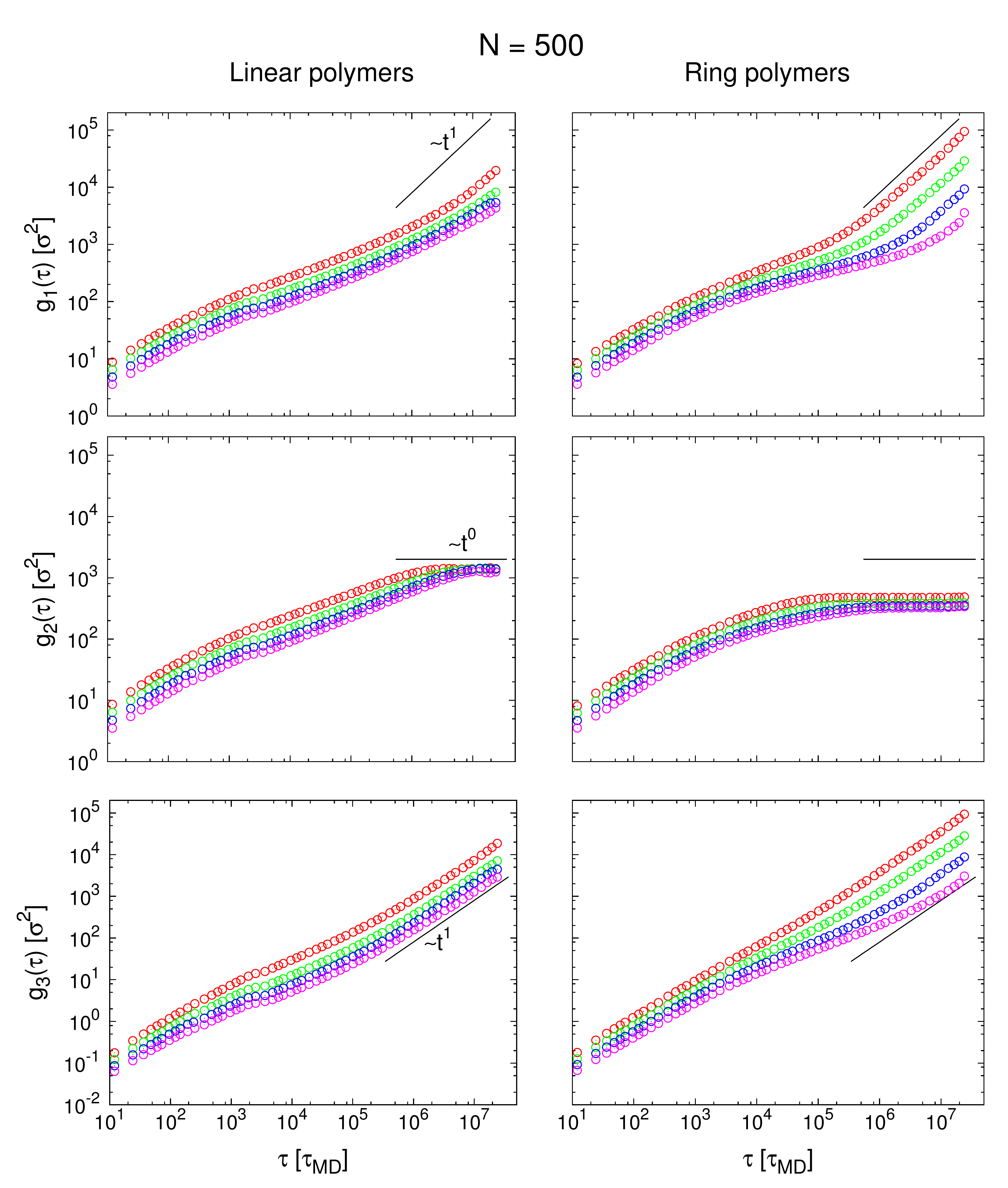}
\end{array}
$$
\caption{
\label{fig:monomerMSD_N500}
Dynamics of chains made of $N=500$ monomers.
Symbols are the same as in Fig.~S\ref{fig:monomerMSD_N250}.
}
\end{figure*}

\begin{figure*}
$$
\begin{array}{c}
\includegraphics[width=5.5in]{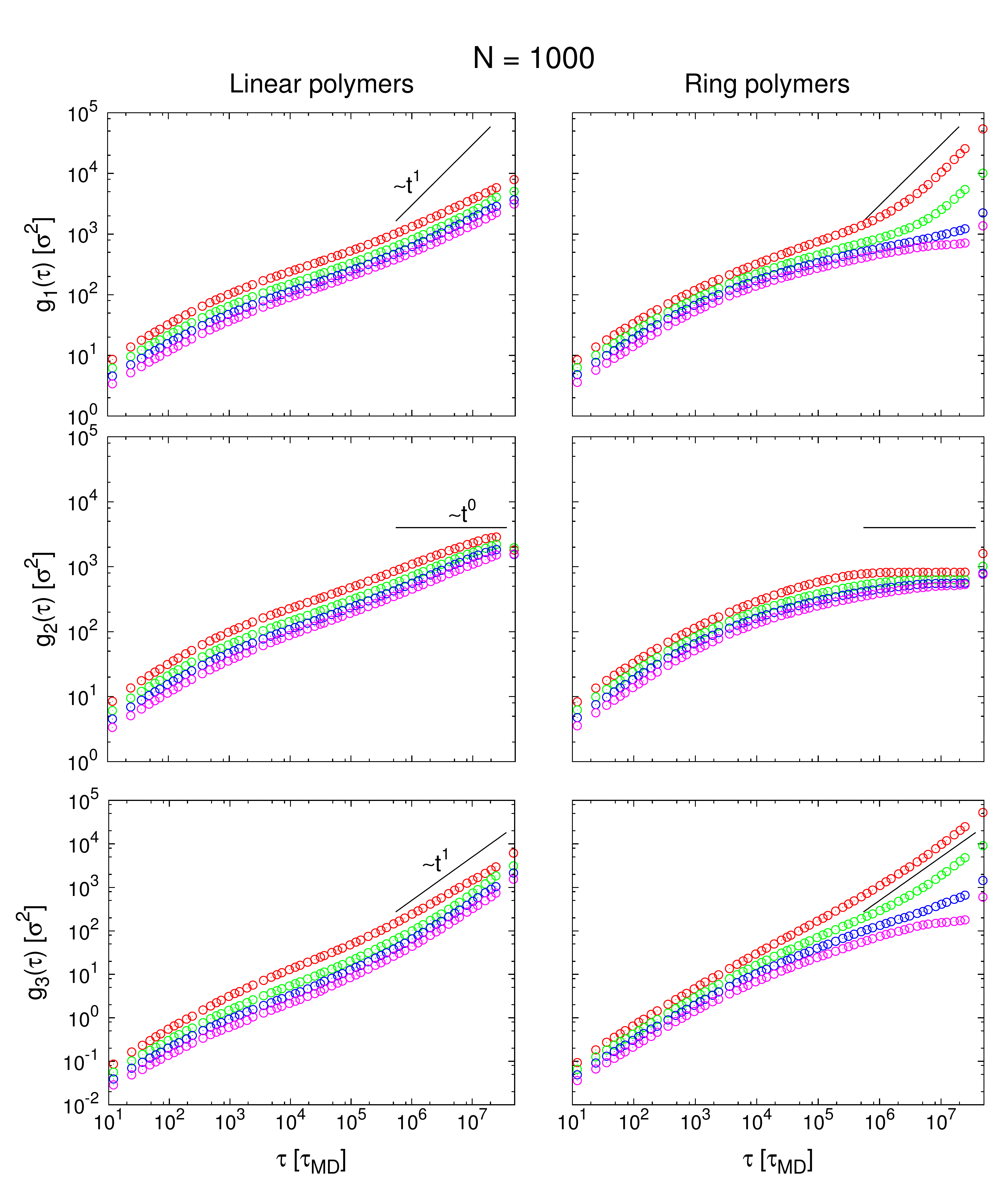}
\end{array}
$$
\caption{
\label{fig:monomerMSD_N1000}
Dynamics of chains made of $N=1000$ monomers.
Symbols are the same as in Fig.~S\ref{fig:monomerMSD_N250}.
}
\end{figure*}


\begin{thebibliography}{49}
\expandafter\ifx\csname natexlab\endcsname\relax\def\natexlab#1{#1}\fi
\expandafter\ifx\csname bibnamefont\endcsname\relax
  \def\bibnamefont#1{#1}\fi
\expandafter\ifx\csname bibfnamefont\endcsname\relax
  \def\bibfnamefont#1{#1}\fi
\expandafter\ifx\csname citenamefont\endcsname\relax
  \def\citenamefont#1{#1}\fi
\expandafter\ifx\csname url\endcsname\relax
  \def\url#1{\texttt{#1}}\fi
\expandafter\ifx\csname urlprefix\endcsname\relax\def\urlprefix{URL }\fi
\providecommand{\bibinfo}[2]{#2}
\providecommand{\eprint}[2][]{\url{#2}}

\bibitem[{\citenamefont{Flory}(1969)}]{Flory1969}
\bibinfo{author}{\bibfnamefont{P.~J.} \bibnamefont{Flory}},
  \emph{\bibinfo{title}{Statistical Mechanics of Chain Molecules}}
  (\bibinfo{publisher}{Interscience, New York}, \bibinfo{year}{1969}).

\bibitem[{\citenamefont{{De Gennes}}(1979)}]{DeGennes}
\bibinfo{author}{\bibfnamefont{P.-G.} \bibnamefont{{De Gennes}}},
  \emph{\bibinfo{title}{Scaling Concepts in Polymer Physics}}
  (\bibinfo{publisher}{Cornell University Press}, \bibinfo{address}{Ithaca},
  \bibinfo{year}{1979}).

\bibitem[{\citenamefont{Doi and Edwards}(1986)}]{DoiEdwards}
\bibinfo{author}{\bibfnamefont{M.}~\bibnamefont{Doi}} \bibnamefont{and}
  \bibinfo{author}{\bibfnamefont{S.~F.} \bibnamefont{Edwards}},
  \emph{\bibinfo{title}{The Theory of Polymer Dynamics}}
  (\bibinfo{publisher}{Oxford University Press}, \bibinfo{address}{New York},
  \bibinfo{year}{1986}).

\bibitem[{\citenamefont{Rubinstein and Colby}(2003)}]{RubinsteinColby}
\bibinfo{author}{\bibfnamefont{M.}~\bibnamefont{Rubinstein}} \bibnamefont{and}
  \bibinfo{author}{\bibfnamefont{R.~H.} \bibnamefont{Colby}},
  \emph{\bibinfo{title}{Polymer Physics}} (\bibinfo{publisher}{Oxford
  University Press}, \bibinfo{address}{New York}, \bibinfo{year}{2003}).

\bibitem[{\citenamefont{Khokhlov and Nechaev}(1985)}]{KhokhlovNechaev85}
\bibinfo{author}{\bibfnamefont{A.~R.} \bibnamefont{Khokhlov}} \bibnamefont{and}
  \bibinfo{author}{\bibfnamefont{S.~K.} \bibnamefont{Nechaev}},
  \bibinfo{journal}{Phys. Lett.} \textbf{\bibinfo{volume}{112A}},
  \bibinfo{pages}{156} (\bibinfo{year}{1985}).

\bibitem[{\citenamefont{Grosberg et~al.}(1988)\citenamefont{Grosberg, Nechaev,
  and Shakhnovich}}]{grosbergJPhysFrance1988}
\bibinfo{author}{\bibfnamefont{A.~Y.} \bibnamefont{Grosberg}},
  \bibinfo{author}{\bibfnamefont{S.~K.} \bibnamefont{Nechaev}},
  \bibnamefont{and} \bibinfo{author}{\bibfnamefont{E.~I.}
  \bibnamefont{Shakhnovich}}, \bibinfo{journal}{J. Phys. France}
  \textbf{\bibinfo{volume}{49}}, \bibinfo{pages}{2095} (\bibinfo{year}{1988}).

\bibitem[{\citenamefont{Obukhov et~al.}(1994)\citenamefont{Obukhov, Rubinstein,
  and Duke}}]{ORD_PRL1994}
\bibinfo{author}{\bibfnamefont{S.~P.} \bibnamefont{Obukhov}},
  \bibinfo{author}{\bibfnamefont{M.}~\bibnamefont{Rubinstein}},
  \bibnamefont{and} \bibinfo{author}{\bibfnamefont{T.}~\bibnamefont{Duke}},
  \bibinfo{journal}{Phys. Rev. Lett.} \textbf{\bibinfo{volume}{73}},
  \bibinfo{pages}{1263} (\bibinfo{year}{1994}).

\bibitem[{\citenamefont{Brereton and Vilgis}(1995)}]{BreretonVilgis1995}
\bibinfo{author}{\bibfnamefont{M.~G.} \bibnamefont{Brereton}} \bibnamefont{and}
  \bibinfo{author}{\bibfnamefont{T.~A.} \bibnamefont{Vilgis}},
  \bibinfo{journal}{J. Phys. A: Math. Gen.} \textbf{\bibinfo{volume}{28}},
  \bibinfo{pages}{1149} (\bibinfo{year}{1995}).

\bibitem[{\citenamefont{M{\"u}ller et~al.}(1996)\citenamefont{M{\"u}ller,
  Wittmer, and Cates}}]{mullerPRE1996}
\bibinfo{author}{\bibfnamefont{M.}~\bibnamefont{M{\"u}ller}},
  \bibinfo{author}{\bibfnamefont{J.~P.} \bibnamefont{Wittmer}},
  \bibnamefont{and} \bibinfo{author}{\bibfnamefont{M.~E.} \bibnamefont{Cates}},
  \bibinfo{journal}{Phys. Rev. E} \textbf{\bibinfo{volume}{53}},
  \bibinfo{pages}{5063} (\bibinfo{year}{1996}).

\bibitem[{\citenamefont{M{\"u}ller et~al.}(2000)\citenamefont{M{\"u}ller,
  Wittmer, and Cates}}]{mullerPRE2000}
\bibinfo{author}{\bibfnamefont{M.}~\bibnamefont{M{\"u}ller}},
  \bibinfo{author}{\bibfnamefont{J.~P.} \bibnamefont{Wittmer}},
  \bibnamefont{and} \bibinfo{author}{\bibfnamefont{M.~E.} \bibnamefont{Cates}},
  \bibinfo{journal}{Phys. Rev. E} \textbf{\bibinfo{volume}{61}},
  \bibinfo{pages}{4078} (\bibinfo{year}{2000}).

\bibitem[{\citenamefont{Kapnistos et~al.}(2008)\citenamefont{Kapnistos, Lang,
  Vlassopoulos, Pyckhout-Hintzen, Richter, Cho, Chang, and
  Rubinstein}}]{kapnistos2008}
\bibinfo{author}{\bibfnamefont{M.}~\bibnamefont{Kapnistos}},
  \bibinfo{author}{\bibfnamefont{M.}~\bibnamefont{Lang}},
  \bibinfo{author}{\bibfnamefont{D.}~\bibnamefont{Vlassopoulos}},
  \bibinfo{author}{\bibfnamefont{W.}~\bibnamefont{Pyckhout-Hintzen}},
  \bibinfo{author}{\bibfnamefont{D.}~\bibnamefont{Richter}},
  \bibinfo{author}{\bibfnamefont{D.}~\bibnamefont{Cho}},
  \bibinfo{author}{\bibfnamefont{T.}~\bibnamefont{Chang}}, \bibnamefont{and}
  \bibinfo{author}{\bibfnamefont{M.}~\bibnamefont{Rubinstein}},
  \bibinfo{journal}{Nature Materials} \textbf{\bibinfo{volume}{7}},
  \bibinfo{pages}{997} (\bibinfo{year}{2008}).

\bibitem[{\citenamefont{Vettorel et~al.}(2009)\citenamefont{Vettorel, Grosberg,
  and Kremer}}]{Vettorel2009}
\bibinfo{author}{\bibfnamefont{T.}~\bibnamefont{Vettorel}},
  \bibinfo{author}{\bibfnamefont{A.~Y.} \bibnamefont{Grosberg}},
  \bibnamefont{and} \bibinfo{author}{\bibfnamefont{K.}~\bibnamefont{Kremer}},
  \bibinfo{journal}{Phys. Biol.} \textbf{\bibinfo{volume}{6}},
  \bibinfo{pages}{025013} (\bibinfo{year}{2009}).

\bibitem[{\citenamefont{Suzuki et~al.}(2009)\citenamefont{Suzuki, Takano,
  Deguchi, and Matsushita}}]{DeguchiJCP2009}
\bibinfo{author}{\bibfnamefont{J.}~\bibnamefont{Suzuki}},
  \bibinfo{author}{\bibfnamefont{A.}~\bibnamefont{Takano}},
  \bibinfo{author}{\bibfnamefont{T.}~\bibnamefont{Deguchi}}, \bibnamefont{and}
  \bibinfo{author}{\bibfnamefont{Y.}~\bibnamefont{Matsushita}},
  \bibinfo{journal}{J. Chem. Phys.} \textbf{\bibinfo{volume}{131}},
  \bibinfo{pages}{144902} (\bibinfo{year}{2009}).

\bibitem[{\citenamefont{Halverson
  et~al.}(2011{\natexlab{a}})\citenamefont{Halverson, Lee, Grest, Grosberg, and
  Kremer}}]{Halverson2011_1}
\bibinfo{author}{\bibfnamefont{J.~D.} \bibnamefont{Halverson}},
  \bibinfo{author}{\bibfnamefont{W.~B.} \bibnamefont{Lee}},
  \bibinfo{author}{\bibfnamefont{G.~S.} \bibnamefont{Grest}},
  \bibinfo{author}{\bibfnamefont{A.~Y.} \bibnamefont{Grosberg}},
  \bibnamefont{and} \bibinfo{author}{\bibfnamefont{K.}~\bibnamefont{Kremer}},
  \bibinfo{journal}{J. Chem. Phys.} \textbf{\bibinfo{volume}{134}},
  \bibinfo{pages}{204904} (\bibinfo{year}{2011}{\natexlab{a}}).

\bibitem[{\citenamefont{Rosa et~al.}(2011)\citenamefont{Rosa, Orlandini,
  Tubiana, and Micheletti}}]{TubianaRosa2011}
\bibinfo{author}{\bibfnamefont{A.}~\bibnamefont{Rosa}},
  \bibinfo{author}{\bibfnamefont{E.}~\bibnamefont{Orlandini}},
  \bibinfo{author}{\bibfnamefont{L.}~\bibnamefont{Tubiana}}, \bibnamefont{and}
  \bibinfo{author}{\bibfnamefont{C.}~\bibnamefont{Micheletti}},
  \bibinfo{journal}{Macromolecules} \textbf{\bibinfo{volume}{44}},
  \bibinfo{pages}{8668} (\bibinfo{year}{2011}).

\bibitem[{\citenamefont{Halverson et~al.}(2012)\citenamefont{Halverson, Grest,
  Grosberg, and Kremer}}]{HalversonPRL2012}
\bibinfo{author}{\bibfnamefont{J.~D.} \bibnamefont{Halverson}},
  \bibinfo{author}{\bibfnamefont{G.~S.} \bibnamefont{Grest}},
  \bibinfo{author}{\bibfnamefont{A.~Y.} \bibnamefont{Grosberg}},
  \bibnamefont{and} \bibinfo{author}{\bibfnamefont{K.}~\bibnamefont{Kremer}},
  \bibinfo{journal}{Phys. Rev. Lett.} \textbf{\bibinfo{volume}{108}},
  \bibinfo{pages}{038301} (\bibinfo{year}{2012}).

\bibitem[{\citenamefont{Sakaue}(2012)}]{SakauePRL2012}
\bibinfo{author}{\bibfnamefont{T.}~\bibnamefont{Sakaue}},
  \bibinfo{journal}{Phys. Rev. Lett.} \textbf{\bibinfo{volume}{106}},
  \bibinfo{pages}{167802} (\bibinfo{year}{2012}).

\bibitem[{\citenamefont{Grosberg}(2012)}]{Grosberg_PolSciC_2012}
\bibinfo{author}{\bibfnamefont{A.~Y.} \bibnamefont{Grosberg}},
  \bibinfo{journal}{Polym. Sci. Ser. C} \textbf{\bibinfo{volume}{54}},
  \bibinfo{pages}{1} (\bibinfo{year}{2012}).

\bibitem[{\citenamefont{Smrek and Grosberg}(2013)}]{SmrekGrosberg2013}
\bibinfo{author}{\bibfnamefont{J.}~\bibnamefont{Smrek}} \bibnamefont{and}
  \bibinfo{author}{\bibfnamefont{A.~Y.} \bibnamefont{Grosberg}},
  \bibinfo{journal}{Physica A} \textbf{\bibinfo{volume}{392}},
  \bibinfo{pages}{6375} (\bibinfo{year}{2013}).

\bibitem[{\citenamefont{Rosa and Everaers}(2014)}]{RosaEveraersPRL2014}
\bibinfo{author}{\bibfnamefont{A.}~\bibnamefont{Rosa}} \bibnamefont{and}
  \bibinfo{author}{\bibfnamefont{R.}~\bibnamefont{Everaers}},
  \bibinfo{journal}{Phys. Rev. Lett.} \textbf{\bibinfo{volume}{112}},
  \bibinfo{pages}{118302} (\bibinfo{year}{2014}).

\bibitem[{\citenamefont{Grosberg}(2014)}]{GrosbergSoftMatter2014}
\bibinfo{author}{\bibfnamefont{A.~Y.} \bibnamefont{Grosberg}},
  \bibinfo{journal}{Soft Matter} \textbf{\bibinfo{volume}{10}},
  \bibinfo{pages}{560} (\bibinfo{year}{2014}).

\bibitem[{\citenamefont{Shin et~al.}({2014})\citenamefont{Shin, Cherstvy, and
  Metzler}}]{CherstvyMetzler2014}
\bibinfo{author}{\bibfnamefont{J.}~\bibnamefont{Shin}},
  \bibinfo{author}{\bibfnamefont{A.~G.} \bibnamefont{Cherstvy}},
  \bibnamefont{and} \bibinfo{author}{\bibfnamefont{R.}~\bibnamefont{Metzler}},
  \bibinfo{journal}{{New J. Phys.}} \textbf{\bibinfo{volume}{{16}}},
  \bibinfo{pages}{{053047}} (\bibinfo{year}{{2014}}).

\bibitem[{\citenamefont{Narros et~al.}(2014)\citenamefont{Narros, Likos,
  Moreno, and Capone}}]{CaponeLikosSoftMatter2014}
\bibinfo{author}{\bibfnamefont{A.}~\bibnamefont{Narros}},
  \bibinfo{author}{\bibfnamefont{C.~N.} \bibnamefont{Likos}},
  \bibinfo{author}{\bibfnamefont{A.~J.} \bibnamefont{Moreno}},
  \bibnamefont{and} \bibinfo{author}{\bibfnamefont{B.}~\bibnamefont{Capone}},
  \bibinfo{journal}{Soft Matter} \textbf{\bibinfo{volume}{10}},
  \bibinfo{pages}{9601} (\bibinfo{year}{2014}).

\bibitem[{\citenamefont{Michieletto et~al.}(2014)\citenamefont{Michieletto,
  Marenduzzo, Orlandini, Alexander, and Turner}}]{MichielettoTurner2014}
\bibinfo{author}{\bibfnamefont{D.}~\bibnamefont{Michieletto}},
  \bibinfo{author}{\bibfnamefont{D.}~\bibnamefont{Marenduzzo}},
  \bibinfo{author}{\bibfnamefont{E.}~\bibnamefont{Orlandini}},
  \bibinfo{author}{\bibfnamefont{G.~P.} \bibnamefont{Alexander}},
  \bibnamefont{and} \bibinfo{author}{\bibfnamefont{M.~S.}
  \bibnamefont{Turner}}, \bibinfo{journal}{ACS Macro Lett.}
  \textbf{\bibinfo{volume}{3}}, \bibinfo{pages}{255} (\bibinfo{year}{2014}).

\bibitem[{\citenamefont{Kremer and Grest}(1990)}]{KremerGrestJCP1990}
\bibinfo{author}{\bibfnamefont{K.}~\bibnamefont{Kremer}} \bibnamefont{and}
  \bibinfo{author}{\bibfnamefont{G.~S.} \bibnamefont{Grest}},
  \bibinfo{journal}{J. Chem. Phys.} \textbf{\bibinfo{volume}{92}},
  \bibinfo{pages}{5057} (\bibinfo{year}{1990}).

\bibitem[{\citenamefont{Auhl et~al.}(2003)\citenamefont{Auhl, Everaers, Grest,
  Kremer, and Plimpton}}]{AuhlJCP2003}
\bibinfo{author}{\bibfnamefont{R.}~\bibnamefont{Auhl}},
  \bibinfo{author}{\bibfnamefont{R.}~\bibnamefont{Everaers}},
  \bibinfo{author}{\bibfnamefont{G.~S.} \bibnamefont{Grest}},
  \bibinfo{author}{\bibfnamefont{K.}~\bibnamefont{Kremer}}, \bibnamefont{and}
  \bibinfo{author}{\bibfnamefont{S.~J.} \bibnamefont{Plimpton}},
  \bibinfo{journal}{J. Chem. Phys.} \textbf{\bibinfo{volume}{119}},
  \bibinfo{pages}{12718} (\bibinfo{year}{2003}).

\bibitem[{\citenamefont{Everaers and Ejtehadi}(2003)}]{EveraersEjtehadi2003}
\bibinfo{author}{\bibfnamefont{R.}~\bibnamefont{Everaers}} \bibnamefont{and}
  \bibinfo{author}{\bibfnamefont{M.~R.} \bibnamefont{Ejtehadi}},
  \bibinfo{journal}{Phys. Rev. E} \textbf{\bibinfo{volume}{67}},
  \bibinfo{pages}{041710} (\bibinfo{year}{2003}).

\bibitem[{\citenamefont{Plimpton}(1995)}]{lammps}
\bibinfo{author}{\bibfnamefont{S.}~\bibnamefont{Plimpton}},
  \bibinfo{journal}{J. Comp. Phys.} \textbf{\bibinfo{volume}{117}},
  \bibinfo{pages}{1} (\bibinfo{year}{1995}).

\bibitem[{\citenamefont{Rosa and Everaers}(2008)}]{RosaPLOS2008}
\bibinfo{author}{\bibfnamefont{A.}~\bibnamefont{Rosa}} \bibnamefont{and}
  \bibinfo{author}{\bibfnamefont{R.}~\bibnamefont{Everaers}},
  \bibinfo{journal}{{PLoS} Comput. Biol.} \textbf{\bibinfo{volume}{4}},
  \bibinfo{pages}{e1000153} (\bibinfo{year}{2008}).

\bibitem[{\citenamefont{Uchida et~al.}(2008)\citenamefont{Uchida, Grest, and
  Everaers}}]{uchida}
\bibinfo{author}{\bibfnamefont{N.}~\bibnamefont{Uchida}},
  \bibinfo{author}{\bibfnamefont{G.~S.} \bibnamefont{Grest}}, \bibnamefont{and}
  \bibinfo{author}{\bibfnamefont{R.}~\bibnamefont{Everaers}},
  \bibinfo{journal}{J. Chem. Phys.} \textbf{\bibinfo{volume}{128}},
  \bibinfo{pages}{044902} (\bibinfo{year}{2008}).

\bibitem[{Not({\natexlab{a}})}]{NoteOnWLREq}
\bibinfo{note}{It is easy to verify that Eq.~\ref{eq:wlr} gives a very accurate
  description of internal distances in worm-like rings since: (1) $\langle
  R^2(l) \rangle \approx l_K^2$ for $l \ll l_K$ and (2) $\langle R^2(l) \rangle
  \approx \left( \frac{1}{l_K l} + \frac{1}{l_K (L-l)} \right)^{-1}$ at
  intermediate $l$ is the known expression for internal distances of flexible
  rings~\cite{RubinsteinColby}.}

\bibitem[{\citenamefont{Becker et~al.}(2010)\citenamefont{Becker, Rosa, and
  Everaers}}]{BeckerRosaEveraers}
\bibinfo{author}{\bibfnamefont{N.~B.} \bibnamefont{Becker}},
  \bibinfo{author}{\bibfnamefont{A.}~\bibnamefont{Rosa}}, \bibnamefont{and}
  \bibinfo{author}{\bibfnamefont{R.}~\bibnamefont{Everaers}},
  \bibinfo{journal}{Eur. Phys. J. E} \textbf{\bibinfo{volume}{32}},
  \bibinfo{pages}{53} (\bibinfo{year}{2010}).

\bibitem[{\citenamefont{Redner}(1980)}]{Redner1980}
\bibinfo{author}{\bibfnamefont{S.}~\bibnamefont{Redner}}, \bibinfo{journal}{J.
  Phys. A: Math. Gen.} \textbf{\bibinfo{volume}{13}}, \bibinfo{pages}{3525}
  (\bibinfo{year}{1980}).

\bibitem[{\citenamefont{des Cloizeaux and Jannink}(1989)}]{DesCloizBook}
\bibinfo{author}{\bibfnamefont{J.}~\bibnamefont{des Cloizeaux}}
  \bibnamefont{and} \bibinfo{author}{\bibfnamefont{G.}~\bibnamefont{Jannink}},
  \emph{\bibinfo{title}{Polymers in Solution}} (\bibinfo{publisher}{Oxford
  University Press}, \bibinfo{address}{Oxford}, \bibinfo{year}{1989}).

\bibitem[{\citenamefont{Everaers et~al.}(1995)\citenamefont{Everaers, Graham,
  and Zuckermann}}]{EveraersJPA1995}
\bibinfo{author}{\bibfnamefont{R.}~\bibnamefont{Everaers}},
  \bibinfo{author}{\bibfnamefont{I.~S.} \bibnamefont{Graham}},
  \bibnamefont{and} \bibinfo{author}{\bibfnamefont{M.~J.}
  \bibnamefont{Zuckermann}}, \bibinfo{journal}{J. Phys. A: Math. Gen.}
  \textbf{\bibinfo{volume}{28}}, \bibinfo{pages}{1271} (\bibinfo{year}{1995}).

\bibitem[{\citenamefont{Rosa et~al.}(2015)}]{RosaGrosbergEveraers2015}
\bibinfo{author}{\bibfnamefont{A.}~\bibnamefont{Rosa}} \bibnamefont{et~al.},
  \bibinfo{journal}{In preparation}  (\bibinfo{year}{2015}).

\bibitem[{Fit()}]{FitDetailsNote}
\bibinfo{note}{Final estimates of exponents $\theta$ and $t$ are obtained by
  averaging the results of two fits to the RdC function, Eq.~\ref{eq:RdC}. In
  the first fit, we consider the complete set of points for $p(R|l)$, while in
  the second one we exclude a few points close to $R=0$. The two procedures
  give very similar results for $t$ (which describes the large-$R$ behaviour of
  $P(R|l)$), while estimates for $\theta$ are typically noisier due to large
  fluctuations of $P(R|l)$ around $R=0$. The outcomes of the two fits are
  combined together as (average value)$\pm$(statistical error)$\pm$(systematic
  error), where the ``statistical error'' is the combination of the two
  statistical errors from the two fitting procedures while the ``systematic
  error'' is the spread between the single estimates, respectively. Final error
  bars reported in Eq.~\ref{eq:RdCfit} are given by $\sqrt{(\mbox{statistical
  error})^2 + (\mbox{systematic error})^2}$.}

\bibitem[{\citenamefont{Fisher}(1966)}]{FisherSAWShape1966}
\bibinfo{author}{\bibfnamefont{M.~E.} \bibnamefont{Fisher}},
  \bibinfo{journal}{J. Chem. Phys.} \textbf{\bibinfo{volume}{44}},
  \bibinfo{pages}{616} (\bibinfo{year}{1966}).

\bibitem[{\citenamefont{Pincus}(1976)}]{PincusBlob1976}
\bibinfo{author}{\bibfnamefont{P.}~\bibnamefont{Pincus}},
  \bibinfo{journal}{Macromolecules} \textbf{\bibinfo{volume}{9}},
  \bibinfo{pages}{386} (\bibinfo{year}{1976}).

\bibitem[{\citenamefont{Debye and Bueche}(1952)}]{DebyeBuecheJCP1952}
\bibinfo{author}{\bibfnamefont{P.}~\bibnamefont{Debye}} \bibnamefont{and}
  \bibinfo{author}{\bibfnamefont{F.}~\bibnamefont{Bueche}},
  \bibinfo{journal}{J. Chem. Phys.} \textbf{\bibinfo{volume}{20}},
  \bibinfo{pages}{1337} (\bibinfo{year}{1952}).

\bibitem[{\citenamefont{Lieberman-Aiden et~al.}(2009)}]{hic}
\bibinfo{author}{\bibfnamefont{E.}~\bibnamefont{Lieberman-Aiden}}
  \bibnamefont{et~al.}, \bibinfo{journal}{Science}
  \textbf{\bibinfo{volume}{326}}, \bibinfo{pages}{289} (\bibinfo{year}{2009}).

\bibitem[{\citenamefont{Halverson
  et~al.}(2011{\natexlab{b}})\citenamefont{Halverson, Lee, Grest, Grosberg, and
  Kremer}}]{Halverson2011_2}
\bibinfo{author}{\bibfnamefont{J.~D.} \bibnamefont{Halverson}},
  \bibinfo{author}{\bibfnamefont{W.~B.} \bibnamefont{Lee}},
  \bibinfo{author}{\bibfnamefont{G.~S.} \bibnamefont{Grest}},
  \bibinfo{author}{\bibfnamefont{A.~Y.} \bibnamefont{Grosberg}},
  \bibnamefont{and} \bibinfo{author}{\bibfnamefont{K.}~\bibnamefont{Kremer}},
  \bibinfo{journal}{J. Chem. Phys.} \textbf{\bibinfo{volume}{134}},
  \bibinfo{pages}{204905} (\bibinfo{year}{2011}{\natexlab{b}}).

\bibitem[{\citenamefont{{De Gennes}}(1971)}]{deGennes71}
\bibinfo{author}{\bibfnamefont{P.-G.} \bibnamefont{{De Gennes}}},
  \bibinfo{journal}{J. Chem. Phys.} \textbf{\bibinfo{volume}{55}},
  \bibinfo{pages}{572} (\bibinfo{year}{1971}).

\bibitem[{\citenamefont{Tseng et~al.}(2004)\citenamefont{Tseng, Lee, Kole,
  Jiang, and Wirtz}}]{TsengWirtz2004}
\bibinfo{author}{\bibfnamefont{Y.}~\bibnamefont{Tseng}},
  \bibinfo{author}{\bibfnamefont{J.~S.~H.} \bibnamefont{Lee}},
  \bibinfo{author}{\bibfnamefont{T.~P.} \bibnamefont{Kole}},
  \bibinfo{author}{\bibfnamefont{I.}~\bibnamefont{Jiang}}, \bibnamefont{and}
  \bibinfo{author}{\bibfnamefont{D.}~\bibnamefont{Wirtz}}, \bibinfo{journal}{J.
  Cell Sci.} \textbf{\bibinfo{volume}{117}}, \bibinfo{pages}{2159}
  (\bibinfo{year}{2004}).

\bibitem[{\citenamefont{Wirtz}(2009)}]{WirtzReview}
\bibinfo{author}{\bibfnamefont{D.}~\bibnamefont{Wirtz}},
  \bibinfo{journal}{Annu. Rev. Biophys.} \textbf{\bibinfo{volume}{38}},
  \bibinfo{pages}{301} (\bibinfo{year}{2009}).

\bibitem[{Not({\natexlab{b}})}]{NoteOnMeshSize}
\bibinfo{note}{The mesh size of the solution is taken of the order of the tube
  diameter~\cite{DoiEdwards} $\sim \sqrt{\langle R_g^2(L_e) \rangle} =
  \sqrt{l_K L_e / 6}$, where $L_e$ is the entanglement length of the solution
  (see the discussion in section~\ref{sec:Results}).}

\bibitem[{Not({\natexlab{c}})}]{NoteOnBoundaryConditions}
\bibinfo{note}{The reader may notice that the effective cross-diameter of the
  colloid particle is given by the sum of its nominal diameter $d$ and the
  monomer diameter $\sigma$. Moreover, due to pure repulsion between colloid
  particles and monomer particles the traditional~\cite{DoiEdwards} geometric
  factor ``$3\pi$'' valid for sticky boundary conditions is substituted by the
  more correct ``$2\pi$'' valid for slip boundary
  conditions~\cite{OuldKaddourLevesque_PRE63_2000} which apply here.}

\bibitem[{\citenamefont{Hihara et~al.}(2012)}]{maeshimaCellRep2012}
\bibinfo{author}{\bibfnamefont{S.}~\bibnamefont{Hihara}} \bibnamefont{et~al.},
  \bibinfo{journal}{Cell Rep.} \textbf{\bibinfo{volume}{2}},
  \bibinfo{pages}{1645} (\bibinfo{year}{2012}).

\bibitem[{\citenamefont{Ould-Kaddour and
  Levesque}(2000)}]{OuldKaddourLevesque_PRE63_2000}
\bibinfo{author}{\bibfnamefont{F.}~\bibnamefont{Ould-Kaddour}}
  \bibnamefont{and} \bibinfo{author}{\bibfnamefont{D.}~\bibnamefont{Levesque}},
  \bibinfo{journal}{Phys. Rev. E} \textbf{\bibinfo{volume}{63}},
  \bibinfo{pages}{011205} (\bibinfo{year}{2000}).

\end{thebibliography}

\end{document}